%% file: LSTM_time_delay_SNIa.tex
\documentclass[traditabstract]{aa} 
\usepackage{graphicx}
\usepackage{grffile}
\usepackage{siunitx}
\usepackage{txfonts}
\usepackage{graphicx}
\usepackage{caption}
\usepackage{natbib}
\usepackage{subfigure} 
\usepackage{hyperref}
\usepackage[utf8]{inputenc}
\usepackage{placeins}
\usepackage{scrextend}
\usepackage{rotating} 
\usepackage{multirow} 
\usepackage{booktabs} 
\usepackage{tablefootnote}
\DeclareUnicodeCharacter{2212}{\textendash}

\newcommand{\slstm}{LSTM-FCNN}

\input{Time_delay_measurement_macro}

\begin{document}

   \title{HOLISMOKES - XII. Time-delay Measurements of Strongly Lensed Type Ia Supernovae using a Long Short-Term Memory Network}

  \titlerunning{Time-delay measurement of strongly lensed SNe Ia}

   \author{S. Huber\inst{1,2}
                 \and    
          S. H. Suyu\inst{1,2,3}
          }

   \institute{Max-Planck-Institut f\"ur Astrophysik, Karl-Schwarzschild Str.~1, 85748 Garching, Germany\\
              \email{shuber@MPA-Garching.MPG.DE}
         \and 
           Technical University of Munich, TUM School of Natural Sciences, Physics Department, James-Franck-Str. 1, 85748 Garching, Germany
           \and
           Academia Sinica Institute of Astronomy and Astrophysics (ASIAA), 11F of ASMAB, No.~1, Section 4, Roosevelt Road, Taipei 10617, Taiwan
}


 
  \abstract
{Strongly lensed Type Ia supernovae (LSNe Ia) are a promising probe to measure the Hubble constant ($H_0$) directly. To use LSNe Ia for cosmography, a time-delay measurement between the multiple images, a lens-mass model, and a mass reconstruction along the line of sight are required. In this work, we present the machine learning network \slstm\ which is a combination of a Long Short-Term Memory Network (LSTM) and a fully-connected neural network (FCNN). The \slstm\ is designed to measure time delays on a sample of LSNe Ia spanning a broad range of properties, which we expect to find with the upcoming Rubin Observatory Legacy Survey of Space and Time (LSST) and for which follow-up observations are planned. With follow-up observations in $i$ band (cadence of one to three days with a single-epoch $5\sigma$ depth of 24.5 mag), we reach a bias-free delay measurement with a precision around 0.7 days over a large sample of LSNe Ia. The \slstm\ is far more general than previous machine learning approaches such as the Random Forest (RF), where a RF has to be trained for each observational pattern separately, and yet the \slstm\ outperforms the RF by a factor of roughly three. Therefore, the \slstm\ is a very promising approach to achieve robust time delays in LSNe Ia, which is important for a precise and accurate constraint on $H_0$.
}
   {}
   {}
   {}
   {}

   \keywords{Gravitational lensing: strong, micro - supernovae: Type Ia supernova - cosmology: distance scale}

   \maketitle
%

\section{Introduction}

The Hubble constant, $H_0$, is the present-day expansion rate of the universe which anchors the age and scale of our universe. However, there is a significant discrepancy in $H_0$ measurements coming from low redshifts, e.g, using type Ia supernovae (SN Ia) where the extragalactic distance ladder has been calibrated with Cepheids \citep[e.g.,][]{Riess:2018byc,Riess:2019cxk,Riess:2020fzl,Riess+2022,2024ApJ...962L..17R,2024arXiv240104777L,2024arXiv240104776A}, in comparison to high-redshift measurements extrapolated from the cosmic microwave background \citep[CMB;][]{Planck:2018vks} by assuming the flat Lambda cold dark matter ($\Lambda$CDM) cosmological model. Independent investigations significantly lower the tension, e.g., using the tip of the red giant branch to a sample of SNe Ia from the Carnegie-Chicago Hubble Program \citep{Freedman:2019jwv,Freedman_2020,Freedman:2021ahq}. While \cite{DiValentino:2021izs} illustrated that low-redshift probes \citep[e.g.,][]{Birrer:2018vtm,Kourkchi:2020iyz,Denzel:2020zuq,Riess:2020fzl} tend to have higher $H_0$ compared to high-redshift probes \citep[e.g.,][]{Planck:2018vks,SPT-3G:2021wgf,SPT-3G:2021eoc}, it is unclear if the $H_0$ tension can be resolved by unaccounted systematics in some of the measurements or if new physics beyond the successful $\Lambda$CDM model is required, such as curvature $\Omega_k \neq 0$, a dark-energy equation of state parameter $w \neq −1$ (i.e. non-constant dark energy density), a time-varying $w$, or a component of dark energy in the early Universe. To identify the origin of this problem, further independent and precise $H_0$ measurements are required. Different groups tackle this task and methods like surface brightness fluctuations \citep[e.g.,][]{Khetan:2020hmh,Blakeslee:2021rqi}, tailored-expanding-photosphere method of SN II \citep[e.g.,][]{Dessart:2005ax,Vogl:2018ckb,Vogl:2019fhc,Csornyei:2023rpw}, megamasers \citep[e.g.,][]{Pesce:2020xfe}, and gravitational waves \citep[e.g.,][]{LIGOScientific:2017adf,Gayathri:2020mra, PhysRevD.103.043520} have been used to successfully measure $H_0$.

Another very promising approach to measure $H_0$ in a single step, without calibrations of the extragalactic distance ladder, is lensing time-delay cosmography \citep{Refsdal:1964} with strongly lensed supernovae (LSNe). Here a SN is lensed by an intervening galaxy or galaxy cluster into multiple images which appear at different moments in time and therefore enable the measurement of a time delay between the LSN images. In addition to the time delay, a mass model of the lensing galaxy is required as well as a reconstruction of the mass along the line of sight. This method has been applied successfully to strongly lensed quasar systems \citep[e.g.,][]{Bonvin:2018dcc,Birrer:2018vtm,2019MNRAS.490..613S,Rusu:2019xrq,Chen:2019ejq,Wong:2019kwg,Birrer+2020}, by the Time-Delay COSMOgraphy \citep[TDCOSMO;][]{Millon:2019slk} organization, consisting of members of the COSmological MOnitoring of GRAvItational Lenses \citep[COSMOGRAIL;][]{2017Courbin}, the $H_0$ Lenses in COSMOGRAIL's Wellspring \citep[H0LiCOW;][]{Suyu:2016qxx}, the Strong lensing at High Angular Resolution Program \citep[SHARP;][]{Chen:2019ejq}, and the STRong-lensing Insights into the Dark Energy
Survey \citep[STRIDES;][]{DES:2018whv} collaborations. Results from six strongly lensed quasars with well-motivated mass models \citep{Wong:2019kwg} \citep[and also a seventh lens;][]{Shajib:2019toy} are in agreement with results from SNe Ia calibrated with Cepheids, but in 3.1$\sigma$ tension with CMB measurements. A new analysis of seven lensed quasars \citep{Birrer+2020} lowers the tension, where the mass-sheet transformation \citep{Falco:1985,Schneider:2013wga} is only constrained by stellar kinematics, leading to broadened uncertainties which are statistically consistent with the study from \cite{Wong:2019kwg} using physically motivated mass models. Even though LSNe are much rarer than lensed quasars, their advantages are the well-defined light curve shape and that they fade away with time, which enables follow-up observations of the lensing galaxy, especially to measure stellar kinematics \citep{Barnabe2011,2017:Yildirim,Shajib:2018,Yildirim:2019vlv} for breaking model degeneracies, such as the mass-sheet degeneracy. Further, the fading nature of the SN helps to derive a more precise lens model from the multiple images of the SN host galaxy, free of contamination from bright point sources \citep{DingEtal21}. In this work, we focus on strongly lensed type Ia SNe (LSNe Ia) which are even more promising than other lensed SNe, given that SNe Ia are standardizable candles which allow one to determine lensing magnifications and therefore yield an additional way to break degeneracies in the lens mass model \citep{2003MNRAS.338L..25O,2015ApJ...811...70R,Foxley-Marrable:2018dzu,2024arXiv240303264W}, provided that microlensing effects (where the stars of the lensing galaxy can significantly magnify and distort spectra and light curves of individual images \citep{Yahalomi:2017ihe, Huber:2019ljb,PierelRodney+2019,Huber:2020dxc,Weisenbach+2021} can be mitigated. 

To date there are five confirmed LSNe Ia, namely iPTF16geu \citep{Goobar:2016uuf}, SN 2022riv \citep{2022TNSAN.169....1K}, SN Zwicky \citep{2022TNSAN.180....1G,2023AAS...24143207R}, SN H0pe \citep{Polletta:2023vam,2024ApJ...961..171F} and SN Encore (D. Newman \& J. Pierel, private communications), and in addition a Type Ia candidate called SN Requiem \citep{Rodney:2021keu}. However, we expect to find between 500 and 900 LSNe Ia \citep{Quimby:2014,GoldsteinNugent:2017,Goldstein:2017bny,Wojtak:2019hsc,LSSTDarkEnergyScience:2023dbi} with next-generation surveys, such as the upcoming Rubin Observatory Legacy Survey of Space and Time \citep[LSST;][]{Ivezic:2008fe} over the duration of ten years. A more conservative estimate, considering only bright and spatially resolved LSNe Ia from \citet[][hereafter OM10]{Oguri:2010} leads to 40 to 100 systems, of which, assuming sufficient follow-up observation (LSST-like $5\sigma$ depth, 2 day cadence), 10 to 25, depending on the LSST observing strategy, provide accurate and precise time delays \citep{Huber:2019ljb}.  Going one magnitude deeper in follow-up monitoring improves the number of LSNe Ia with well measured time delays by roughly a factor of 1.5 \citep{Huber:2019ljb}. The LSST observing strategy is currently beeing optimized \citep{Lochner:2018, LSSTDarkEnergyScience:2021ryz} which is important to maximize the number of LSNe Ia with well measured time delay. However, many different science cases depend on the LSST observing strategy, leaving only little room for optimization for the science case of LSNe Ia. Therefore it is also very important to improve the time-delay measurement methods, which is the scope of this work.

To measure the time delays between the multiple images of mock LSNe Ia, \cite{Huber:2019ljb} used the free-knot spline estimator from Python Curve Shifting \citep[\texttt{PyCS};][]{2013:Tewesb,Bonvin:2015jia}. Another approach first suggested by \cite{Goldstein:2017bny} uses color curves instead of light curves, in order to minimize the influence of microlensing
\citep{Goldstein:2017bny,
  Huber:2020dxc}. To calculate the microlensing effects on SNe Ia, theoretical models are required, where synthetic observables have been calculated with radiative transfer codes such as \texttt{SEDONA} \citep{Kasen:2006ce} or \texttt{ARTIS} \citep{Kromer:2009ce}. The theoretical models provide the specific intensity and therefore yield a spatial distribution of the radiation which can be combined with magnification factors coming from random source positions in a microlensing map \citep{Vernardos:2014lna,Vernardos:2014yva,Vernardos:2015wta,ChanEtal21} to calculate microlensed spectra, light curves, and color curves. \cite{Goldstein:2017bny} found for the spherical symmetric W7 model \citep{1984:Nomoto} that the color curves are achromatic, meaning almost free of microlensing, up to three rest-frame weeks after explosion.
We have confirmed this result for three other theoretical SN Ia models \citep{Huber:2020dxc}, testing also effects of asymmetries in the SN ejecta, although we have
shown that, in practice, the usage of color curves is limited given that strong features such as peaks are only present in three out of five independent LSST color curves (coming from \textit{ugrizy}), which would require follow-up observations in \textit{rizyJH} for typical source redshifts of LSNe Ia, where it will be especially challenging to get good quality data in the redder bands.

Therefore \cite{Huber:2021iug} investigated the usage of light curves further to measure time delays between the resolved images of LSNe Ia using two simple machine learning techniques: (i) a deep learning fully connected neural network (FCNN) with two hidden layers, and (ii) a random forest \citep[RF;][]{breiman2001random} which is a set of random regression trees. The data for a FCNN or a RF require always the same input structure and size and since LSNe Ia have a variable amount of observed data points, a machine learning network needs to be trained for each observed LSN Ia individually. To include microlensing in the simulations, \cite{Huber:2021iug} trained on four theoretical SN Ia models also used in \cite{Suyu:2020opl,Huber:2020dxc}. While the results on a test set based on the same four theoretical models look very promising for the FCNN and the RF, tests of generalizability, using empirical light curves from the \texttt{SNEMO15} model \citep{Saunders:2018rjn} show that only the RF provides almost bias-free results. For the RF we can achieve an accuracy in the $i$ band within 1\% for a LSNe Ia detected about eight to ten days before peak, if delays are longer than 15 days. The precision for a well sampled LSNe Ia at the median source redshift of 0.77 is around 1.4 days, assuming observations are taken in the $i$ band for a single-epoch $5\sigma$ depth of 24.5 mag with a cadence of 2 days. Another important point is that for typical LSNe Ia source redshifts between 0.55 and 0.99 (16th to 84th percentile of LSNe Ia from OM10), the $i$ band provides the best precision, followed by $r$ and $z$
(or $g$ if $\sourcez \lesssim 0.6$),
where the best performance is achieved if a RF is trained for each filter separately. Further results are that the observational noise is by far the dominant source of uncertainty, although microlensing is not negligible (factor of two in comparison to no microlensing) and that there is no gain in training a single RF for a quad LSNe Ia instead of a separate RF per pair of images, which means treating it like a double system. Furthermore, from the RF we can expect slightly more precise time-delay measurements than with PyCS used by \cite{Huber:2019ljb}.

Another time-delay measurement tool was developed by \cite{PierelRodney+2019}, which uses SN templates from empirical models, but microlensing on LSNe Ia is approximated as wavelength independent\footnote{Lensing is always achromatic, meaning independent of wavelength; however the wavelength dependent spatial distribution of the SN radiation will lead to chromatic effects.}, in contrast to \cite{Huber:2021iug}, where microlensing is calculated from the specific intensity profiles of theoretical models. Further, \cite{Bag_2021}, \cite{Denissenya:2022eds}, and Bag et al. (in preparation) show that time-delay measurement of unresolved LSNe in LSST might be possible using machine learning if they are classified correctly as LSNe.

The main disadvantage of the machine learning models trained by \cite{Huber:2021iug} is that the same input structure and amount of data points is required, meaning that a model needs to be trained for each observation of a LSNe Ia individually. Therefore the goal of this work is to develop a much more general method that can handle LSNe Ia of almost any kind and ideally also outperforms the RF. Specifically, we use a Long Short-Term Memory Network (LSTM), which is well suited to handle time-dependent problems \citep{hochreiter1997long}, in combination with a FCNN, which we refer to as \slstm. In this work, we investigate only LSNe Ia, although the techniques can be also applied to other types of SNe. Applications to other SN types require, however, mock microlensed light curves of these SNe, which are beyond the scope of this paper to produce. The number of LSNe II is slightly higher than the number of LSNe Ia, which makes them also promising for time-delay cosmography. A study on LSNe IIP using spectra and color curves for the time-delay measurement was done by \cite{Bayer:2021ugw}.

This paper is organized as follows. In Sect. \ref{sec: Data set for machine learning} we explain the production of our data set for our machine learning model, which will be explained in Sect. \ref{sec:Machine learning technique used in this work}, before we present our results in Sect. \ref{sec: Results}. In Sect. \ref{sec:specific applications of LSTM network} we apply the \slstm\ to specific configurations and compare results to previous ones presented in \cite{Huber:2021iug}. Finally, we summarize in Sect. \ref{sec:Summary}. Magnitudes in this paper are in the AB system. To calculate distances we assume a flat $\Lambda$CDM cosmology with $H_0 = 72 \, \mathrm{km} \, \mathrm{s}^{−1} \, \mathrm{Mpc}^{−1}$, $\Omega_\mathrm{r} = 0$ (for radiation) and $\Omega_\mathrm{m} = 0.26$ (for matter).

\section{Data set for machine learning}
\label{sec: Data set for machine learning}

In this Section we explain the production of the data set to train, validate, and test our LSTM model. The calculation of microlensed light curves with observational noise is described in Sect. \ref{sec:Microlensed light curves with observational noise} and the creation of the data set used in this work is explained in Sect. \ref{sec:Data set for a LSTM Network}. 

\subsection{Microlensed light curves with observational noise}
\label{sec:Microlensed light curves with observational noise}
The calculation of microlensed light curves is described in detail in \cite{Huber:2019ljb} and the calculation including the observational noise, especially for different moon phases can be found in \cite{Huber:2021iug}. In the following we summarize the most important points.

To calculate light curves including microlensing, a theoretical SN model is required.  We use four of these models to increase the variety: (1) the W7 model \citep{1984:Nomoto}, which is a parametrized 1D deflagration model of a Chandrasekhar-mass (Ch) carbon-oxygen (CO) white dwarf (WD), (2) the delayed detonation N100 model \citep{Seitenzahl:2013} of a Ch CO WD, (3) a sub-Chandrasekhar (sub-Ch) detonation model of a CO WD with $1.06$ solar masses \citep{Sim:2010}, and (4) a merger model of two CO WDs with solar masses of 0.9 and 1.1 \citep{Pakmor:2012}. The theoretical models are used in combination with magnification maps from \texttt{GERLUMPH} \citep{Vernardos:2014lna,Vernardos:2014yva,Vernardos:2015wta} to calculate the microlensed observed flux via:

\begin{equation}
F_{\lambda,\mathrm{o}}(t)=\frac{1}{\lum^2(1+\sourcez)}\int \dd x \int \dd y \, I_{\lambda,\mathrm{e}}(t,p(x,y)) \, \mu(x,y),
\label{eq: microlensed flux}
\end{equation}
with the luminosity distance to the source $\lum$, the source redshift $\sourcez$, the magnification factor $\mu(x,y)$, and the emitted specific intensity at the source
plane $I_{\lambda,\mathrm{e}}(t,p)$, which is a function of wavelength $\lambda$, time since explosion $t$,
and impact parameter $p$, i.e. the projected distance from the ejecta
center, where spherical symmetry is assumed following \cite{Huber:2019ljb, Huber:2020dxc}. The magnification maps which provide the magnification factor $\mu(x,y)$ as a function of the source positions $(x,y)$ depend on three main parameters, the convergence $\kappa$, the shear $\gamma$, and the smooth matter
fraction $s=1-\kappa_*/\kappa$, where $\kappa_*$ is defined as the convergence of
the stellar component. Further, these maps also depend on the source redshift $\sourcez$ and lens redshift
$\lensz$, which set the Einstein Radius $\Rein$ and thus the characteristic size of the map. The microlensing magnification maps are created following \citet{ChanEtal21}, where the maps in \cite{Huber:2021iug} and in this work use a Salpeter initial mass function (IMF) with a mean mass of the microlenses of $0.35 M_\odot$ and a resolution of 20000 $\times$ 20000 pixels with a total size of 20 $\Rein$ $\times$ 20 $\Rein$.

From the observed microlensed flux from Eq. (\ref{eq: microlensed flux}) we obtain the microlensed light curves in AB magnitudes via
\begin{equation}
\scalebox{1.13}{$
m_\mathrm{AB,X}(t) = -2.5 \log_{10} \left(\frac{\int \dd\lambda \, \lambda \, S_\mathrm{X}(\lambda) \, F_{\lambda,\mathrm{o}}(t)  }{\int \dd \lambda \, S_\mathrm{X}(\lambda) \, c \, / \lambda} \times \si{\frac{\square\cm}{\erg}} \right)  - 48.6$}
\label{eq: microlensed light curves for ab magnitudes}
\end{equation}
\citep{Bessel:2012}. Further, $c$ is the speed of light and the transmission function for the filter X is $S_\mathrm{X}(\lambda)$. From Eq. (\ref{eq: microlensed light curves for ab magnitudes}) we can then calculate magnitudes with observational uncertainties via

\begin{equation}
m_\mathrm{X}(t) = m_{\mathrm{AB,X}}(t) + r_\mathrm{norm} \sigma_{\mathrm{X}}(t),
\label{eq:noise realization random mag including error LSST science book}
\end{equation}
where $r_\mathrm{norm}$ is a random number following a Gaussian distribution with zero mean and a standard deviation of one. The $1\sigma$ standard deviation of $m_\mathrm{X}(t)$ is $\sigma_{\mathrm{X}}(t)$, which depends mainly on $m_\mathrm{AB,X}(t)$ relative to the $5\sigma$ depth of the filter X (see \cite{2009:LSSTscience}, Sect. 3.5, p.67, \cite{Huber:2021iug}, and Appendix \ref{sec:Appendix LSST uncertainty}). In this work we assume as in \cite{Huber:2021iug} that follow-up observations will be triggered for the LSNe Ia detected by LSST to get better quality light curves with a mean single-epoch LSST-like $5\sigma$ depth plus one, i.e., 24.5 mag for the $i$ band, which is feasible even with a 2-meter telescope \citep{Huber:2019ljb}. Further, we assume as in \cite{Huber:2021iug} a time-varying $5\sigma$ depth to account for the moon phase.

\subsection{Data set for a Long Short-Term Memory Network}
\label{sec:Data set for a LSTM Network}

In this section we describe the production of the data set for the LSTM network, which
is more general than the RF from \cite{Huber:2021iug} 
since LSTM's architecture can handle most LSNe Ia expected from OM10, without the requirement of specific training for a particular observational pattern. Our training set is composed of a sample of 200,000 LSNe Ia and the validation and test set each have a size of 20,000. A single sample $S$ contains all observed data points for each of the two images of a LSNe Ia, where a single data point $k$ of image $j$ is represented by

\begin{equation}
S_{jk} := \{t_{jk}, m_{jk}, \sigma_{jk} \},
\label{eq:data structure time, mag, unc}
\end{equation}
where $t_{jk}$ is the time when the magnitude $m_{jk}$ of image $j$ is observed with the $1\sigma$ uncertainty
of $\sigma_{jk}$. Since we investigate only the $i$ band in this work, we drop the filter notation as introduced in Eq. (\ref{eq:noise realization random mag including error LSST science book}).  Therefore, the single sample $S$, containing a single LSNe Ia, can be written as
\begin{equation}
S := (S^*_1; S^*_2) := (S_{11}, S_{12}, S_{13}, ..., S_{1N_\mathrm{sl,1}}; S_{21}, S_{22}, ..., S_{2N_\mathrm{sl,2}}),
\label{eq:sample}
\end{equation}
where $N_\mathrm{sl,1}$ and $N_\mathrm{sl,2}$ is the number of observed data points as in Eq. (\ref{eq:data structure time, mag, unc}) for image one and two, which we also refer to as sequence length. The $S^*_1$, respectively $S^*_2$, contains the sequence of observed data points for image one, respectively, image two. If we consider multiple samples, we label them by $l$ leading to $S_l$ $=(S^*_{1,l},S^*_{2,l})$ with $N_{\mathrm{sl,1},l}$ and $N_{\mathrm{sl,2},l}$ for the sequence length of image one and two of the $l$-th sample.

\begin{figure*}[htbp]
\centering
\subfigure{\includegraphics[width=0.45\textwidth]{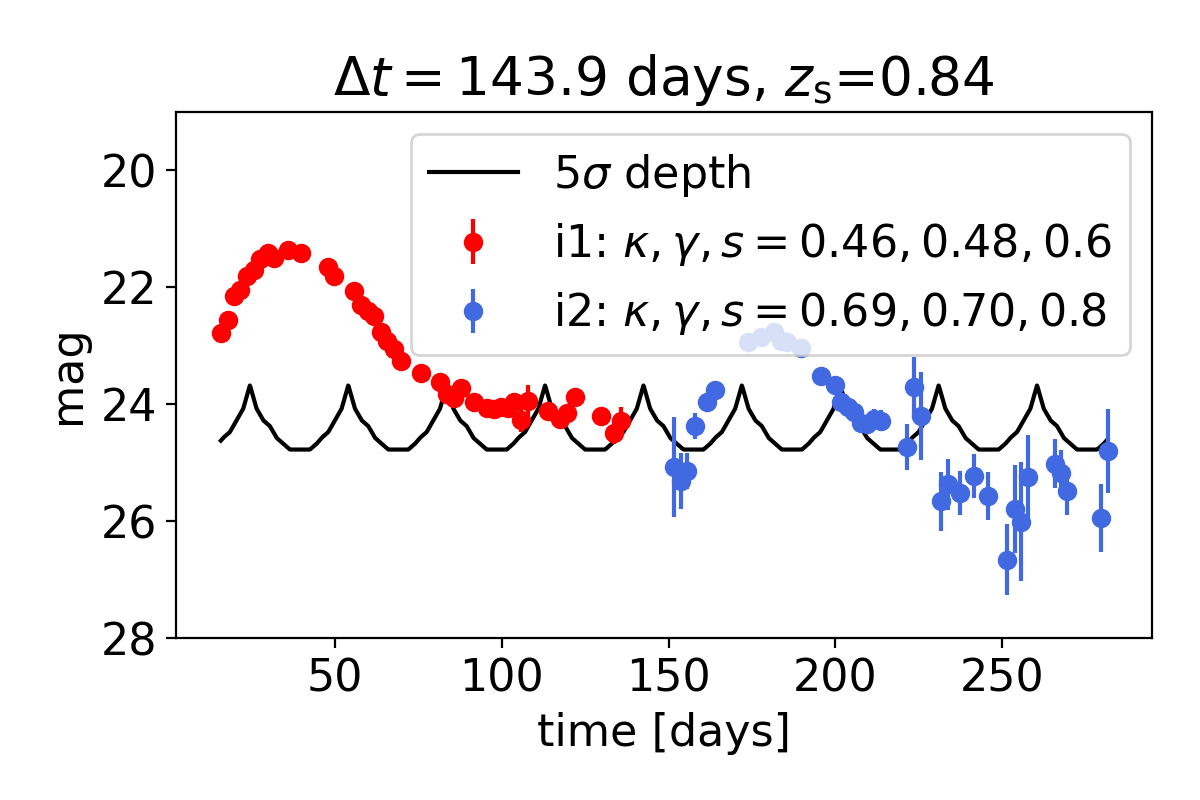}}
\subfigure{\includegraphics[width=0.45\textwidth]{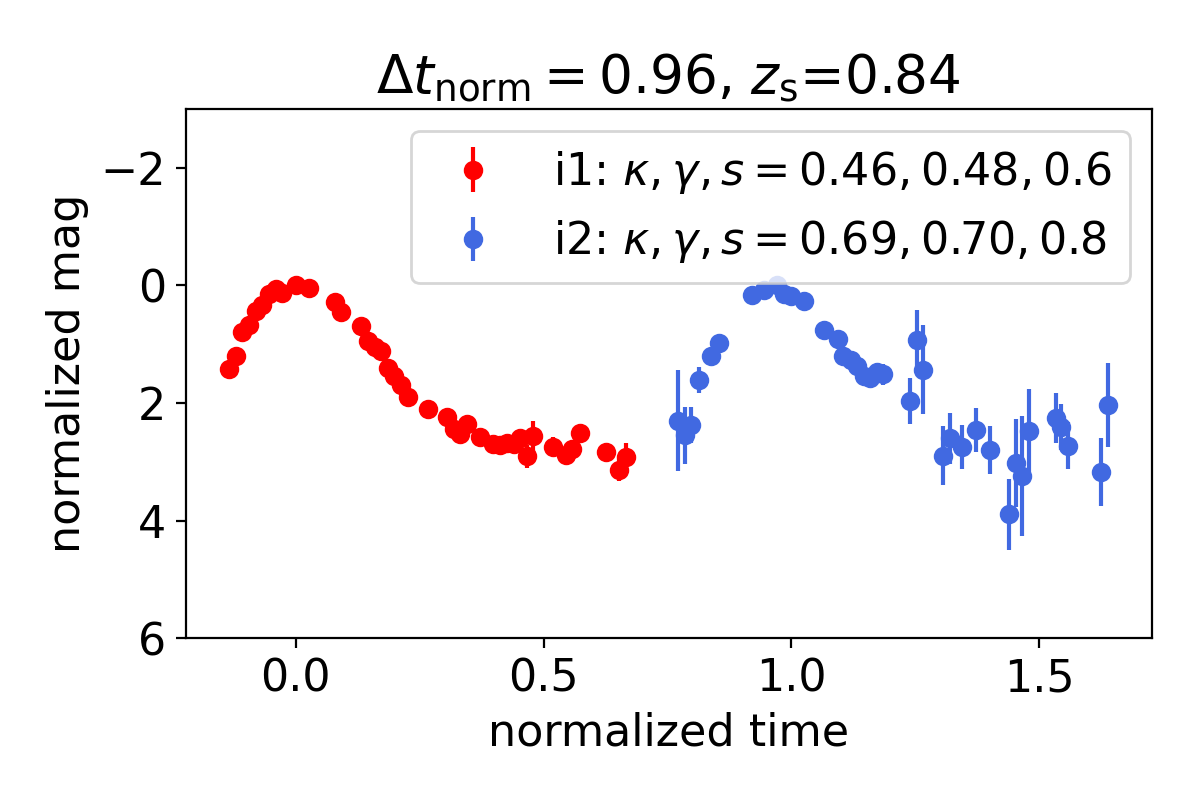}}
\subfigure{\includegraphics[width=0.45\textwidth]{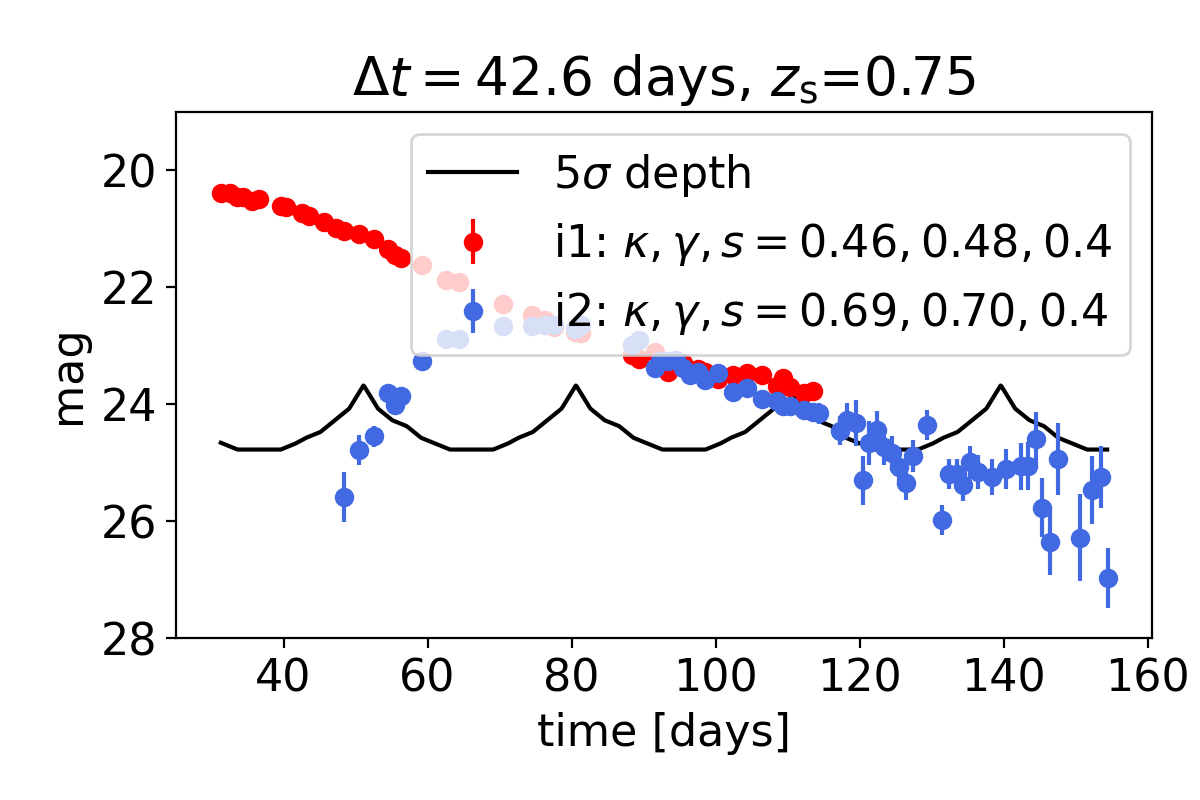}}
\subfigure{\includegraphics[width=0.45\textwidth]{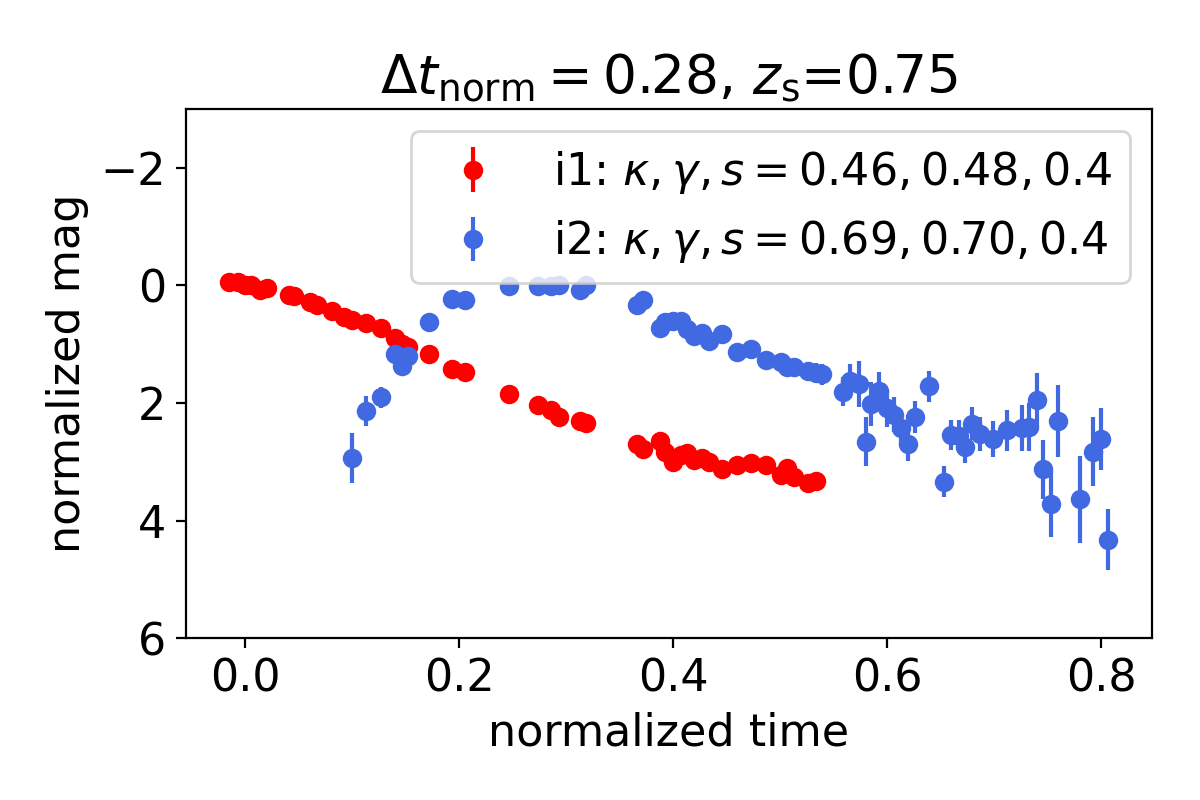}}
\subfigure{\includegraphics[width=0.45\textwidth]{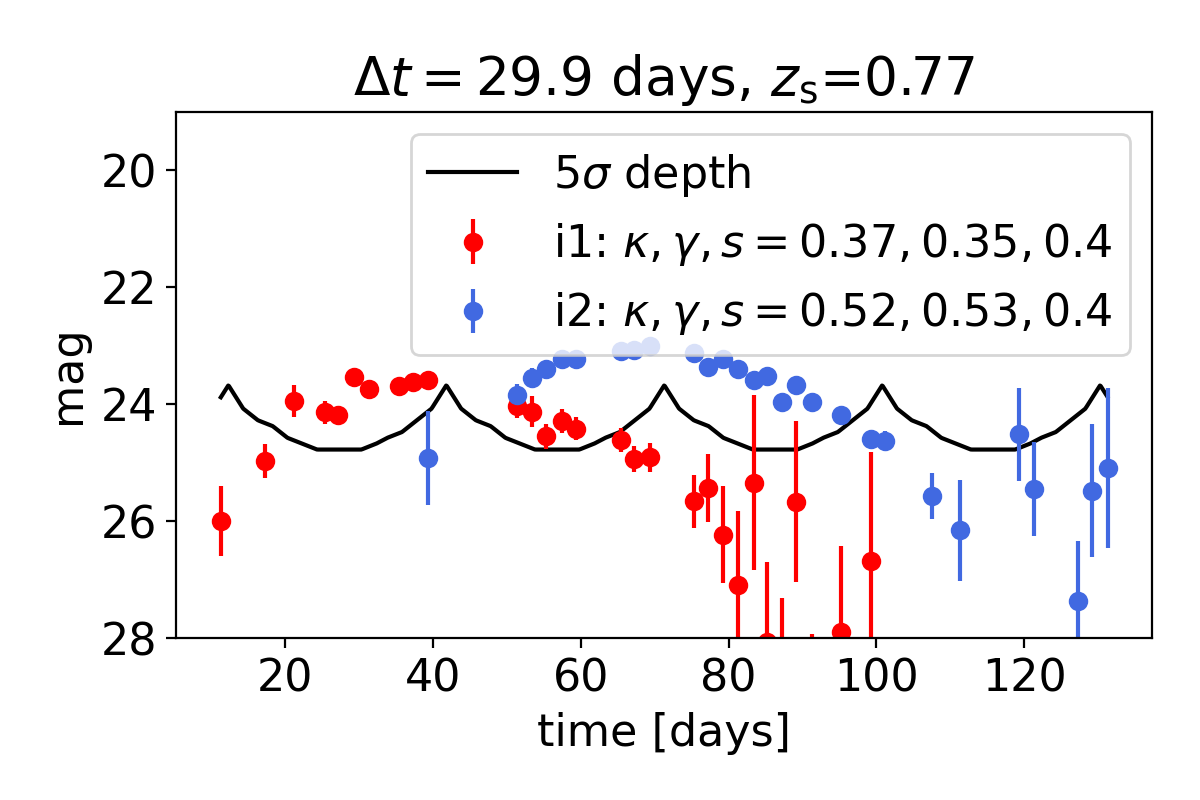}}
\subfigure{\includegraphics[width=0.45\textwidth]{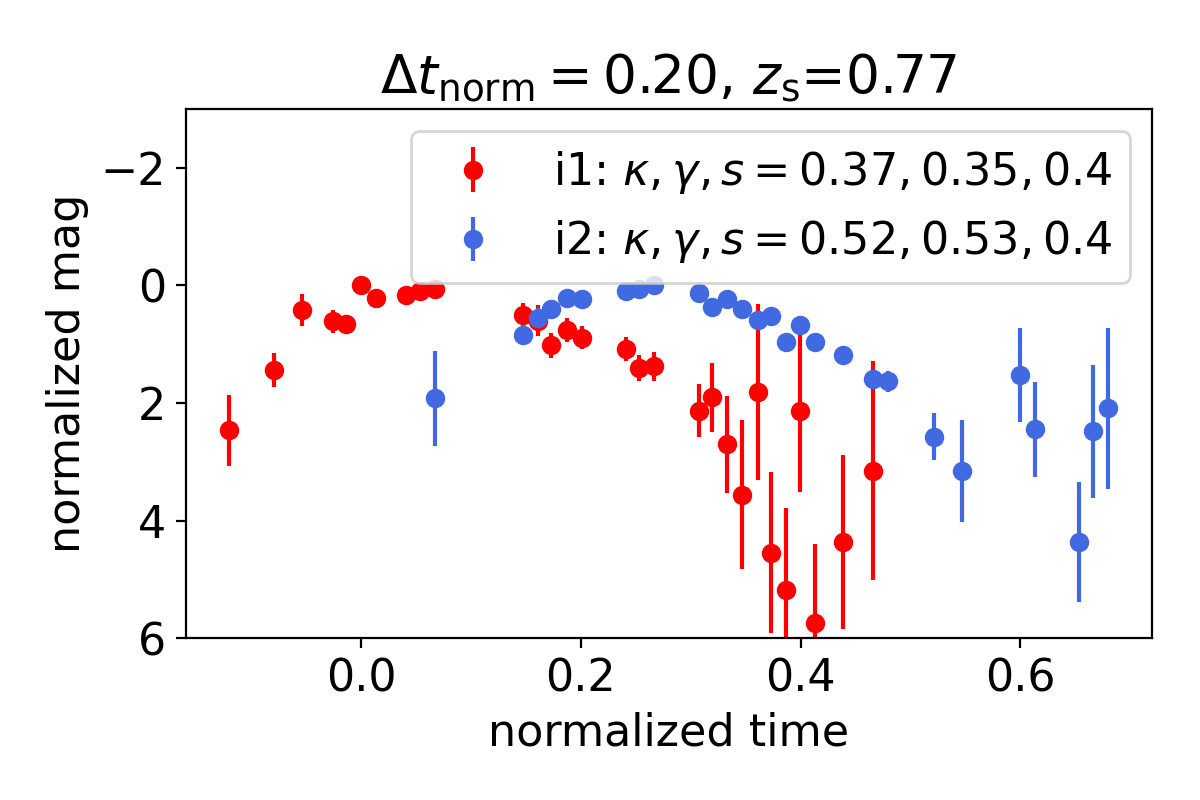}}
\subfigure{\includegraphics[width=0.45\textwidth]{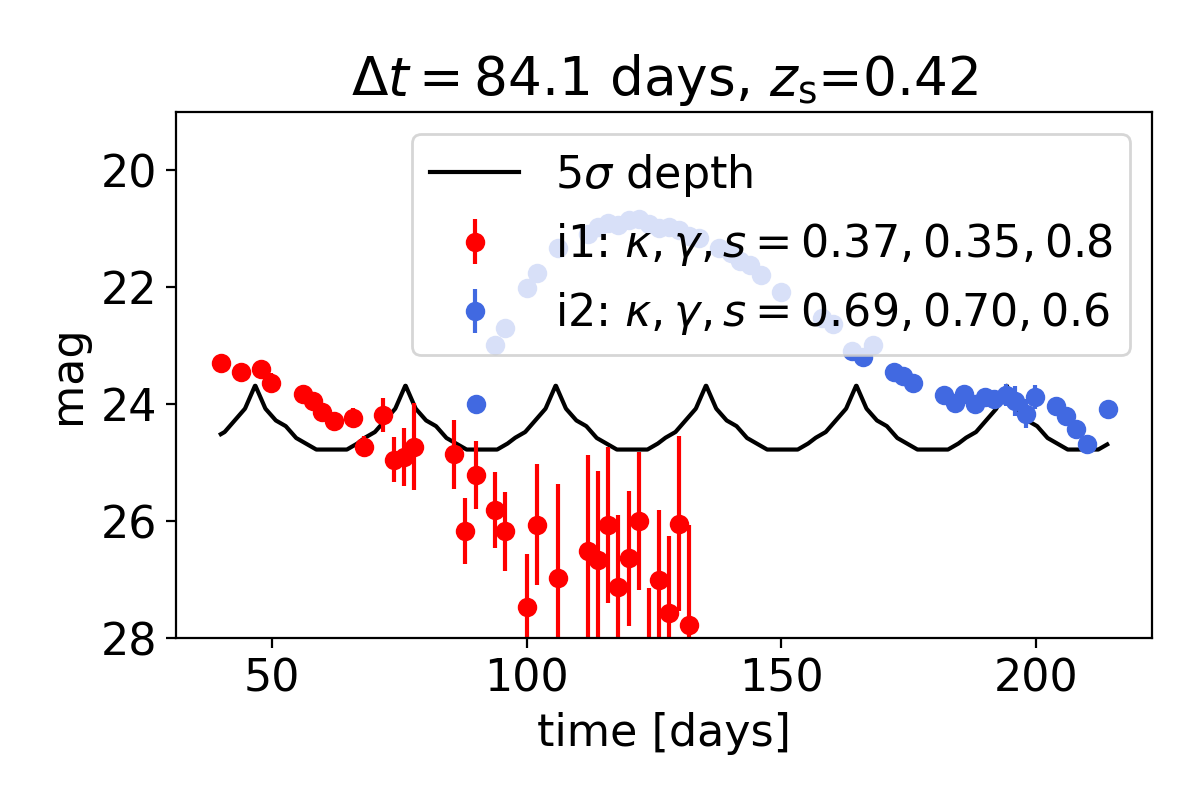}}
\subfigure{\includegraphics[width=0.45\textwidth]{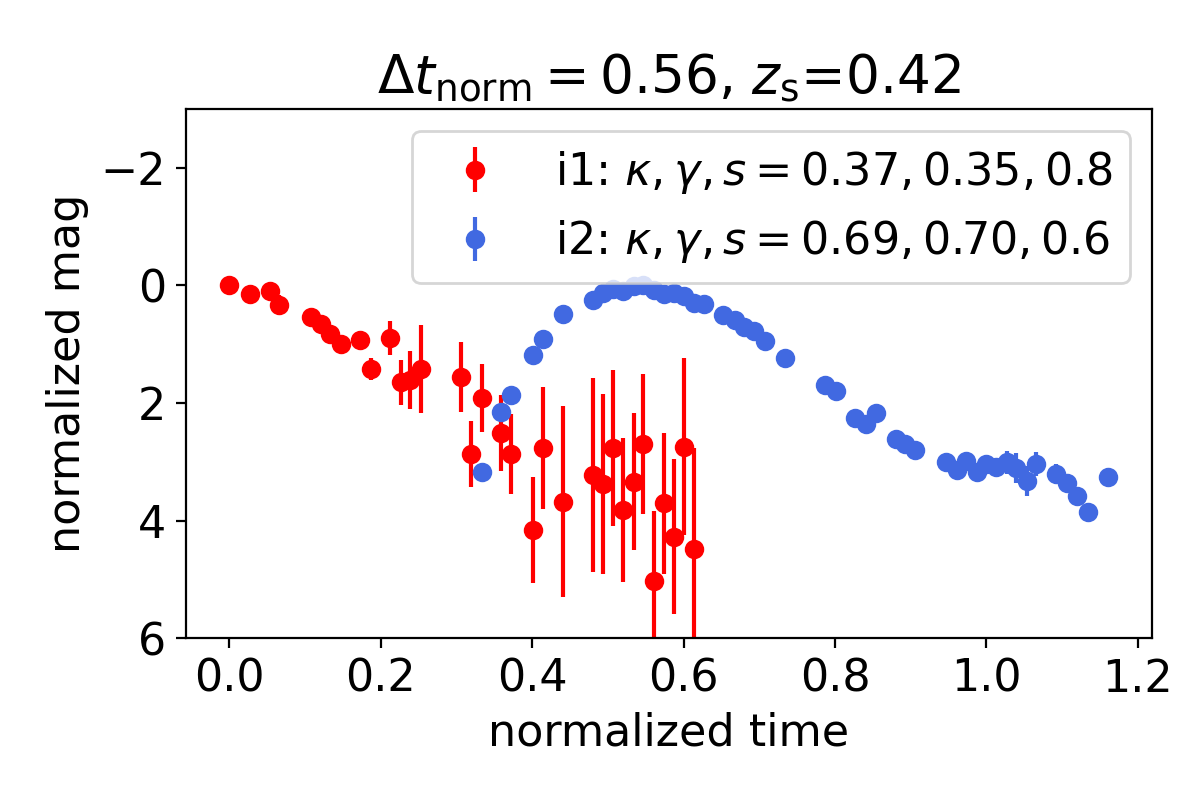}}
\caption{Four random samples of our training set, comparing data without normalization (left panels) and data with normalization (right panels) that will be used for our machine learning network. The red and blue data points are the $i$-band light curves for the first and the second, respectively, lensed SN image. The black curves in the left panels show the $i$-band 5$\sigma$ limiting depth, which oscillates with the moon phase.  The normalization procedure to obtain the right panels from the left panels is described in Sec.~\ref{sec:Data set for a LSTM Network}.}
\label{fig:random_examples_training_data}
\end{figure*}

A random set of LSNe Ia systems used for our training process is shown in Fig. \ref{fig:random_examples_training_data}, where the left four panels show the data without normalization. The right four panels show the same four samples with normalization, which will be used in our machine learning network, where we use all data points (as in Eq. (\ref{eq:data structure time, mag, unc})) from the two images as input to predict the time delay $\Delta t$ between these images. 
The applied normalization is very important such that the input values for the LSTM network have always the same order of magnitude. Therefore, we normalize the magnitude of both images with respect to the peak of the image under consideration. Given that very noisy data points can be brighter than the peak we only consider 25\% of the data points with the lowest uncertainty to find the peak (minimum in magnitude). We then subtract the peak magnitude from each data point in the corresponding light curve, such that the peak has a magnitude value of zero. Furthermore, we also normalize the time scale $t$, which in our case is the time after explosion, by adding a constant offset to $t$ such that the offset $t=0$ corresponds to the peak determined for the first image.  We then divide the time delay $\Delta t$ and time scale $t$ by 150 days, which is the maximum of the time-delays under consideration, such that the output of the LSTM network is restricted to values between 0 and 1.

In the following, we explain how such a random data set is created, where we summarize the variety of lensed SNe that we consider in Table \ref{tab:random parameters for LSTM training}. To simulate light curves for LSNe Ia, we use the four theoretical models described in Sect. \ref{sec:Microlensed light curves with observational noise} and the empirical \texttt{SNEMO15} model \citep{Saunders:2018rjn}, which contains 171 SN Ia light curves. To ensure that we have light curves not used in the training process to test the generalizability, we randomly split the whole \texttt{SNEMO15} data set into three subsets: (1) \texttt{SNEMO15-T} consisting of 87 curves for training, (2) \texttt{SNEMO15-V} consisting of 42 curves for validation, and (3) \texttt{SNEMO15-E} consisting of 42 curves for testing. Our training, validation, and test set is then composed of 50\% light curves from theoretical SN models and 50\% of empirical light curves from the corresponding subset of \texttt{SNEMO15} light curves. Furthermore, we construct a second test set to investigate the generalizability, namely the \texttt{SNEMO15}-only test set, based only on the 42 light curves in \texttt{SNEMO15-E}. A summary of all data sets is listed in Table \ref{tab:train validation test set for theo plus empirical in training process}. 
\begin{table}
\caption[Summary of assumptions for the LSTM data set]{Summary of our inputs for the LSTM data set used for training, validation, and testing.}
\begin{tabular}{l|ccccc}
theoretical SN Ia models & W7, N100, Sub-Ch, merger \\[0.07cm]
empirical SN Ia model & \texttt{SNEMO15} \\[0.07cm]
\multirow{2}{*}{light curve coverage} & $\sim$3.4 to 70.0 rest-frame days\\ &after explosion \\[0.07cm]
absolute  mag (rest-frame $u$) & $-17.9$ to $-19.7$ \\[0.07cm]
absolute  mag (rest-frame $g$) & $-18.7$ to $-19.7$ \\[0.07cm]
time delay & 0 to 150 days \\[0.07cm]
detection of first image & at latest 5 days after peak  & \\[0.07cm]
cadence & 1,2,3 days $\pm$ 4 hours \\[0.07cm]
missing data points & 0 to 50 \% \\[0.07cm]
band / filter & LSST $i$ band \tablefootnote{see e.g.~Fig.~1 in \cite{Huber:2020dxc}}  \\[0.07cm]
mean single-epoch $5\sigma$ depth & 24.5 mag \\[0.07cm]
noise calculation & see Appendix \ref{sec:Appendix LSST uncertainty} \\[0.07cm]
max. $\sigma_{i}(t_k)$ & 2 mag \\[0.07cm]
moon phase & random \\[0.07cm]
source redshift $\sourcez$ & 0.17 to 1.34 following OM10 \\[0.2cm]
\hline
\\
source and lens redshift & (0.77, 0.30), (0.97, 0.47),  \\[0.07cm] 
($\sourcez, \lensz$) only for microlensing 
&  (0.53, 0.13),  (0.94, 0.20), \\[0.07cm] & (0.60, 0.42) \\[0.2cm]
\hline
\\
\multirow{3}{*}{$(\kappa, \gamma)$} & (0.37, 0.35), (0.27, 0.24), \\[0.07cm]
& (0.46, 0.48), (0.69, 0.7), \\[0.07cm]
& (0.90, 0.90), (0.52, 0.53) \\[0.2cm]
\hline
\\
$s$ & 0.4, 0.6, 0.8
\end{tabular}
\centering
\label{tab:random parameters for LSTM training}
\end{table}\begin{table}
  \caption[Description of the training, validation, test, and \texttt{SNEMO15}-only test set for the \slstm]{Description of the training, validation, test, and \texttt{SNEMO15}-only test set used in this work. The \texttt{SNEMO15} light curves are split randomly into three distinct subsets with no overlap in light curves: \texttt{SNEMO15-T} for training, \texttt{SNEMO15-V} for validation and \texttt{SNEMO15-E} for testing.  This ensures that light curves used in one process (e.g., validation) are not used in another process (e.g., training or testing).
}
\begin{tabular}{l|l}
\multirow{4}{*}{Training set} & 50\% light curves from theoretical \\ & SN Ia models \\ [0.07cm]   &  + \\ [0.07cm]  & 50\% light curves from \texttt{SNEMO15-T}  \\[0.2cm] \\
\hline
\\
\multirow{4}{*}{Validation set} & 50\% light curves from theoretical \\
& SN Ia models \\ [0.07cm]   & + \\ [0.07cm]  & 50\% light curves from  \texttt{SNEMO15-V} \\[0.2cm] \\
\hline
\\
\multirow{4}{*}{Test set} & 50\% light curves from theoretical \\ & SN Ia models \\ [0.07cm]   & + \\ [0.07cm]  & 50\% light curves from  \texttt{SNEMO15-E} \\[0.2cm]\\
\hline
\\
\multirow{1}{*}{\texttt{SNEMO15}-only test set} & 100\% light curves  from  \texttt{SNEMO15-E}  \\[0.2cm]
\end{tabular}
\centering
\label{tab:train validation test set for theo plus empirical in training process}
\end{table}

\begin{figure}[htbp]
\centering
\subfigure{\includegraphics[width=0.45\textwidth]{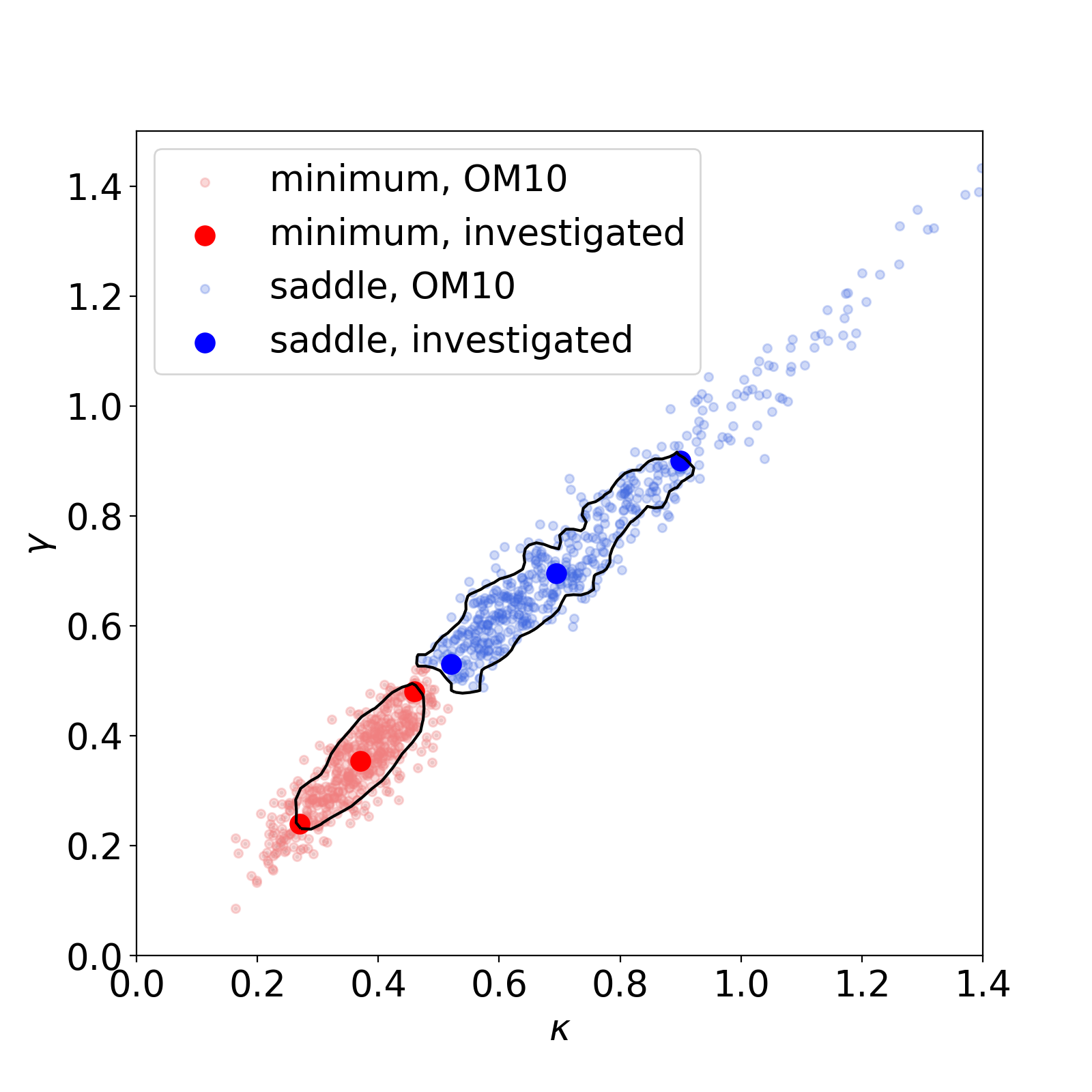}}
\subfigure{\includegraphics[width=0.45\textwidth]{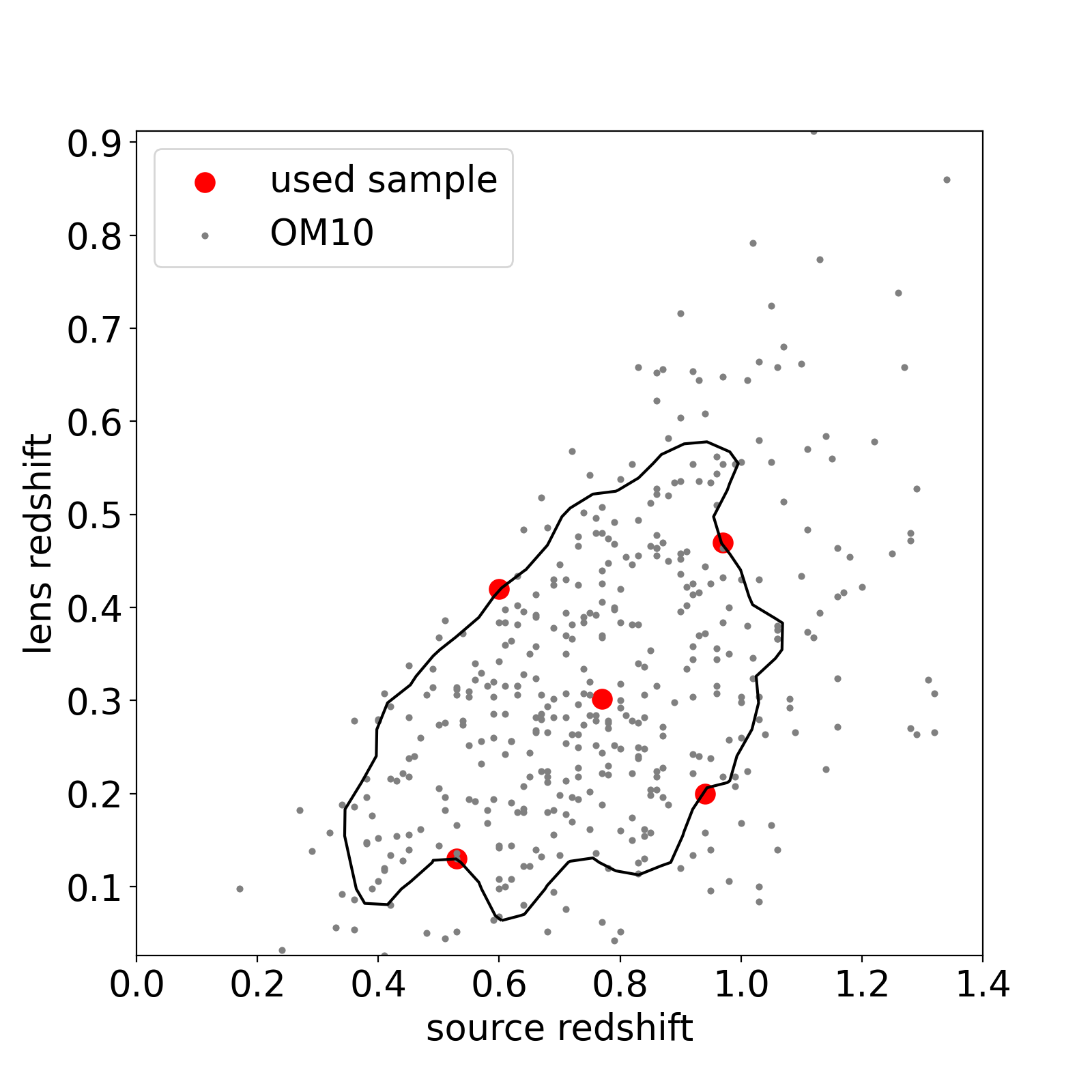}}
\caption{Convergence $\kappa$ and shear $\gamma$ (upper panel), and source redshift $\sourcez$ and lens redshift $\lensz$ (lower panel) used for the calculation of microlensed light curves drawn from the OM10 catalog. The black lines represent 
contours that enclose 68\% of the LSN Ia systems in the OM10 catalog.}
\label{fig:kappa_gamma_source_lens_OM10}
\end{figure}

For each theoretical model, we calculate for a given LSN Ia image, 10000 microlensed light curves from the corresponding microlensing map of the LSN Ia image (set by $\kappa,\gamma,s,\sourcez,\lensz)$, as described in Sect. \ref{sec:Microlensed light curves with observational noise}. Since the calculation of microlensed SN Ia light curves is very time consuming, we only investigate a subset of microlensing magnification maps, while covering a broad range of properties of the LSN Ia image. The upper panel of Fig.~\ref{fig:kappa_gamma_source_lens_OM10} shows the distribution of ($\kappa$, $\gamma$) values from the OM10 catalog and the six pairs used in this work, which are composed of the median and two systems on the $1\sigma$ contours, taken separately for the lensing image types ``minimum'' and ``saddle''. Furthermore, we assume three typical $s$ values of 0.4, 0.6, and 0.8 similar to \cite{Huber:2019ljb,Huber:2020dxc,Huber:2021iug} and five different pairs of ($\sourcez$, $\lensz$) as shown in the lower panel of Fig. \ref{fig:kappa_gamma_source_lens_OM10} also based on the OM10 catalog. Therefore we investigate in total $6 \times 3 \times 5 = 90$ different microlensing magnification maps where we calculate for each map 10000 random source positions for all four theoretical models providing $90 \times 10000 \times 4 = 3600000$ microlensed SN Ia light curves, which is more than sufficient given that \cite{Huber:2021iug} found that microlensing is not the dominant source of uncertainty. 

\begin{figure}[htbp]
\centering
\includegraphics[width=0.48\textwidth]{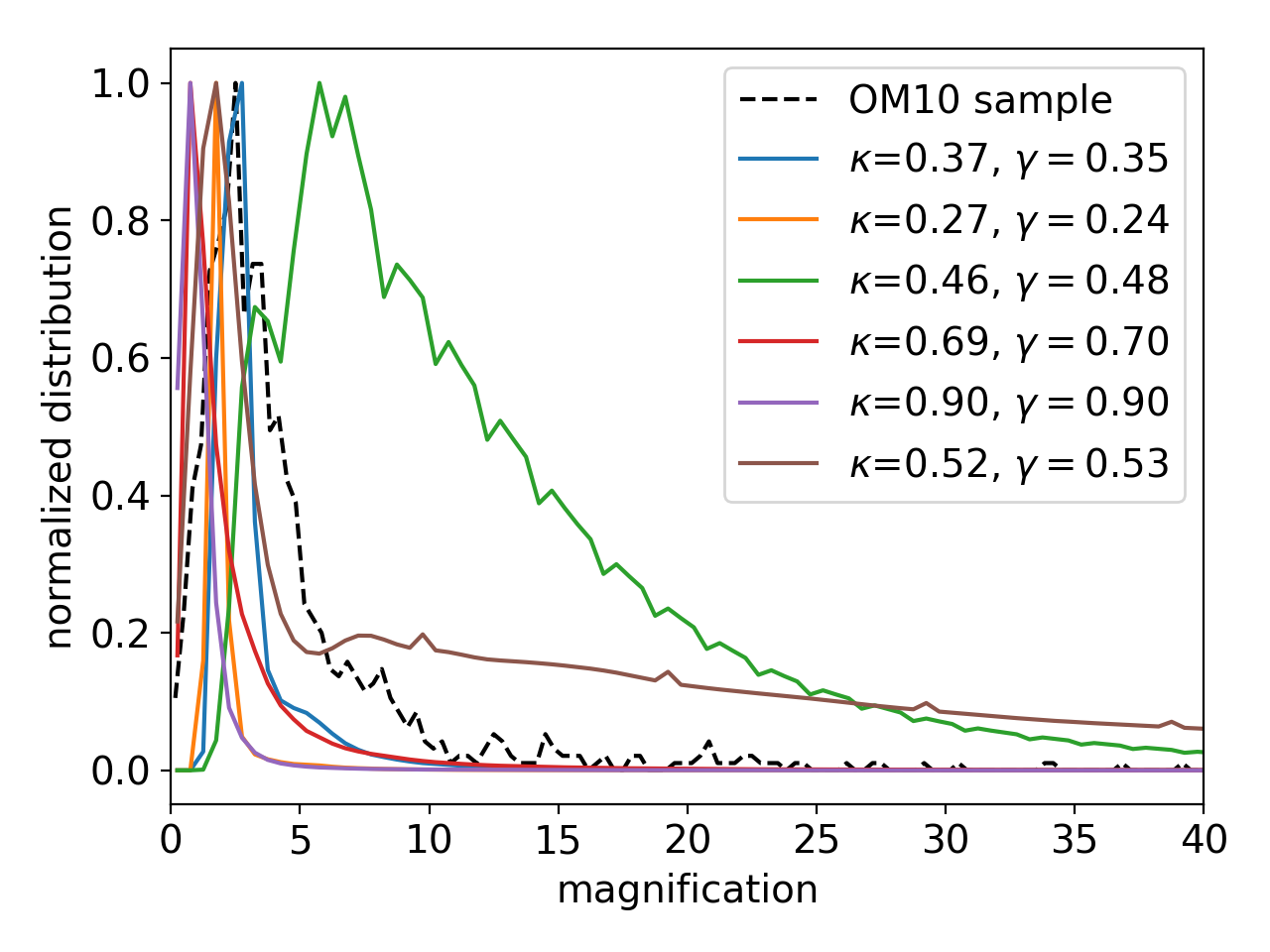}
\caption{Distribution of magnification factors from the six microlensed magnification maps under investigation in this work (solid colored), in comparison to all macro-magnification factors from all LSNe Ia from the OM10 catalog (dashed black). Microlensing magnification maps investigated in this work cover a much broader range of potential magnification factors in comparison to the macrolensing only case from the whole OM10 sample.}
\label{fig:magnification_distribution_micro_vs_OM10}
\end{figure}

For the empirical \texttt{SNEMO15} model, no specific intensity profiles are available, and therefore the calculation of microlensed light curves as described in Sect. \ref{sec:Microlensed light curves with observational noise} is not possible. To nonetheless include microlensing effects, we use the theoretical models where we subtract from a randomly selected microlensed light curve (which include macrolensing plus microlensing) the macrolensed light curve, assuming $\mu_{\rm macro} = 1/((1-\kappa)^2 - \gamma^2)$, to get only the contribution from microlensing which we then add to a \texttt{SNEMO15} light curve. With the six pairs of $\kappa$ and $\gamma$, we only cover a small part of potential macrolensing factors; however, from Fig. \ref{fig:magnification_distribution_micro_vs_OM10}, we see that the microlensed light curves from these six pairs span a much broader range of potential magnification factors as we would get from macrolensing only, from the whole OM10 sample.

The source redshift and the lens redshift together set the scale of the microlensing map, and affect only weakly the microlensing effect on SN Ia light curves \citep{Huber:2020dxc}. However, the source redshift also determines the observed light curve shapes and brightness in a certain band. Since \cite{Huber:2021iug} showed that in terms of noise contributions, the observational noise is far more important than microlensing, and the five source redshifts shown in Fig. \ref{fig:kappa_gamma_source_lens_OM10} are sufficient for the microlensing calculation, but not for the various brightness and light curve shapes coming from the OM10 sample, which include $\sourcez$ between 0.17 and 1.34. Therefore, we draw randomly pairs of ($z_\mathrm{s, OM10}$, $z_\mathrm{d, OM10}$) from the OM10 catalog and we pick from the lower panel of Fig.~\ref{fig:kappa_gamma_source_lens_OM10} the pair ($z_\mathrm{s,micro}$, $z_\mathrm{d,micro}$) from our used sample (in red) which is closest. We then rescale the pre-calculated microlensed flux $F_{\lambda,\mathrm{o},z_\mathrm{s,micro}}$ for the source redshift $z_\mathrm{s,micro}$ to $z_\mathrm{s, OM10}$, the random source redshift of interest:

\begin{equation}
F_{\lambda,\mathrm{o},z_\mathrm{s,OM10}} = F_{\lambda,\mathrm{o},z_\mathrm{s,micro}} \frac{D_\mathrm{lum,z_\mathrm{s,micro}}^2 (1+z_\mathrm{s,micro})}{D_\mathrm{lum,z_\mathrm{s,OM10}}^2 (1+z_\mathrm{s,OM10})}.
\label{eq:redshift corrected flux}
\end{equation}
Further, we rescale as well the wavelength
\begin{equation}
\lambda_{z_\mathrm{s,OM10}} = \frac{\lambda_{z_\mathrm{s,micro}}}{1+z_\mathrm{s,micro}} (1+z_\mathrm{s,OM10})
\end{equation}
and time since explosion
\begin{equation}
t_{z_\mathrm{s,OM10}} = \frac{t_{z_\mathrm{s,micro}}}{1+z_\mathrm{s,micro}} (1+z_\mathrm{s,OM10}).
\label{eq:redshift corrected time}
\end{equation}
With Eq. (\ref{eq: microlensed light curves for ab magnitudes}) we can then calculate the microlensed light curves for $z_\mathrm{s,OM10}$, providing the correct light curve shape and brightness but only approximated microlensing contributions.

For our data set, we further consider a range of the follow-up strategy, namely we pick for each simulated sample randomly a cadence of 1, 2, or 3 days with a random uniform scatter of $\pm4$ hours, given that we will not always observe at the exact same time in a night. Further, we assume a depth 1 mag deeper than the mean LSST single-epoch $5\sigma$ depth, which provides a good compromise between required time on a hypothetical 2m telescope on the one hand and accurate/precise time delays in LSN Ia systems on the other hand \citep{Huber:2019ljb}. In reality we will not be able to observe with a regular cadence because of, e.g., bad weather and therefore we delete randomly between 0 to 50\% of the times follow-up observations are assumed. Furthermore, we limit the discoveries of the first image of a LSN Ia to a detection not later than 5 days after peak, which is the aim of our lens finding approaches. For simplicity, we only train the network in the $i$ band, given that this provides by far the most precise time-delay measurement \citep{Huber:2021iug}. From Eq. (\ref{eq:noise realization random mag including error LSST science book}) we see that extremely high uncertainty values are possible, which occur only rarely in practice and can be easily removed from the data. Therefore, we will only take data into account where the standard deviation $\sigma_{ik}$ of an observed data point $m_{ik}$ is smaller than 2 mag.

\begin{figure}[htbp]
\centering
\subfigure{\includegraphics[width=0.4\textwidth]{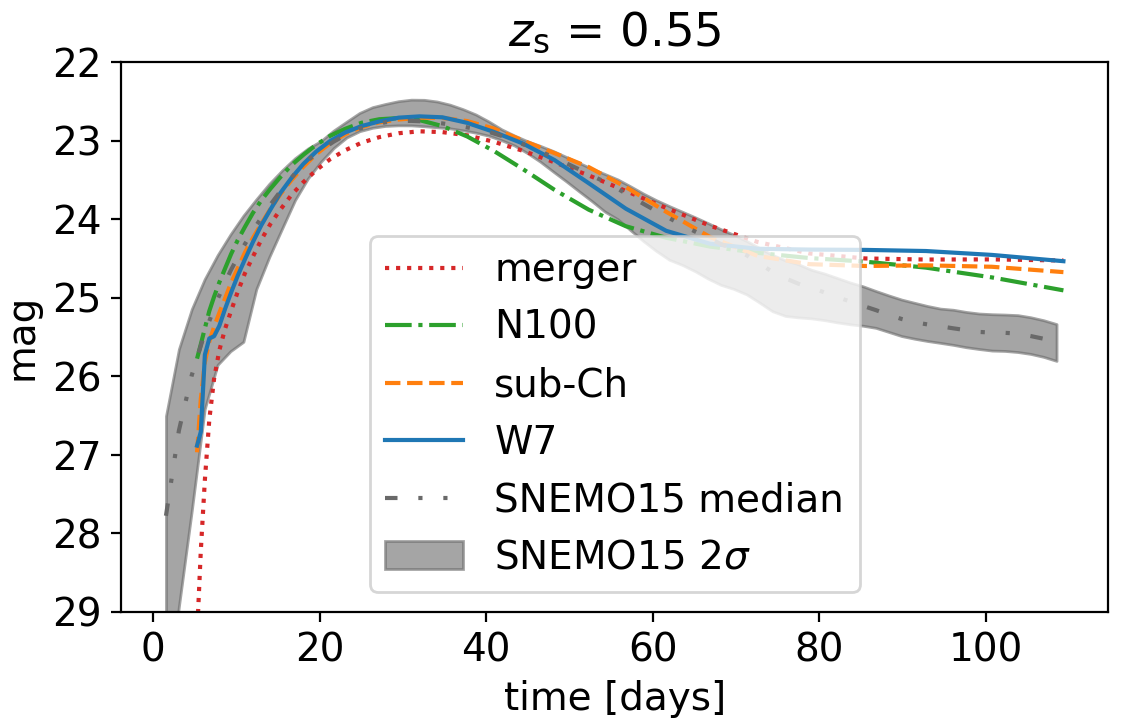}}
\subfigure{\includegraphics[width=0.4\textwidth]{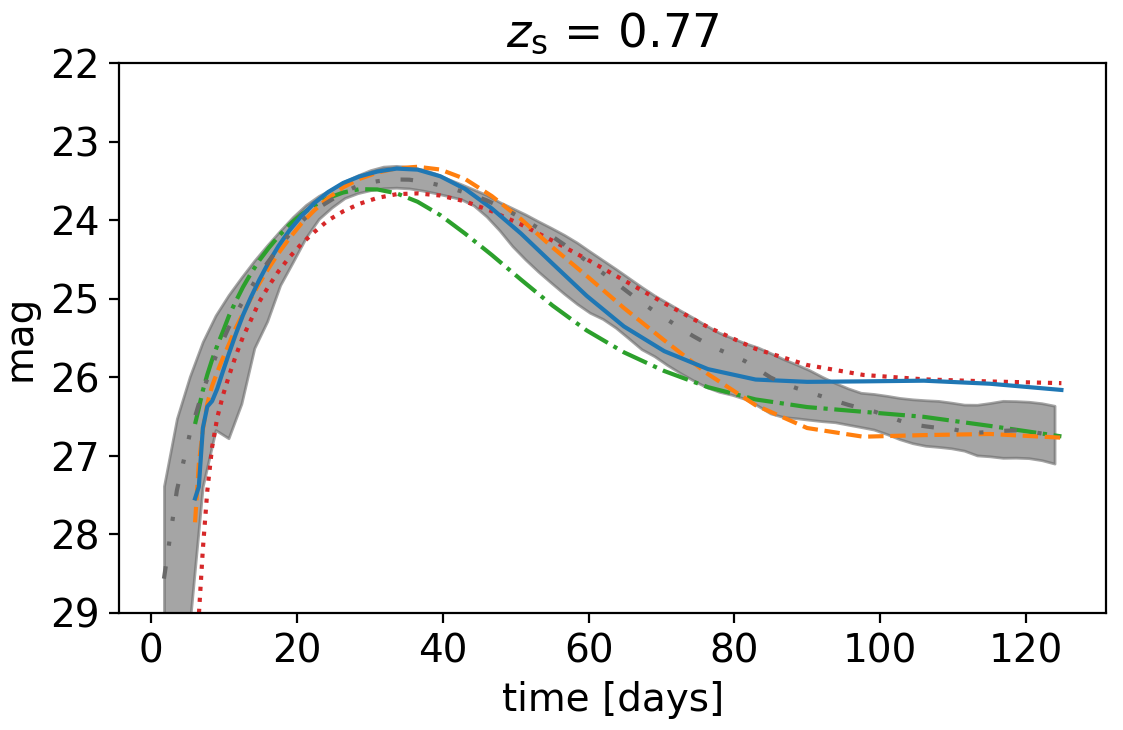}}
\subfigure{\includegraphics[width=0.4\textwidth]{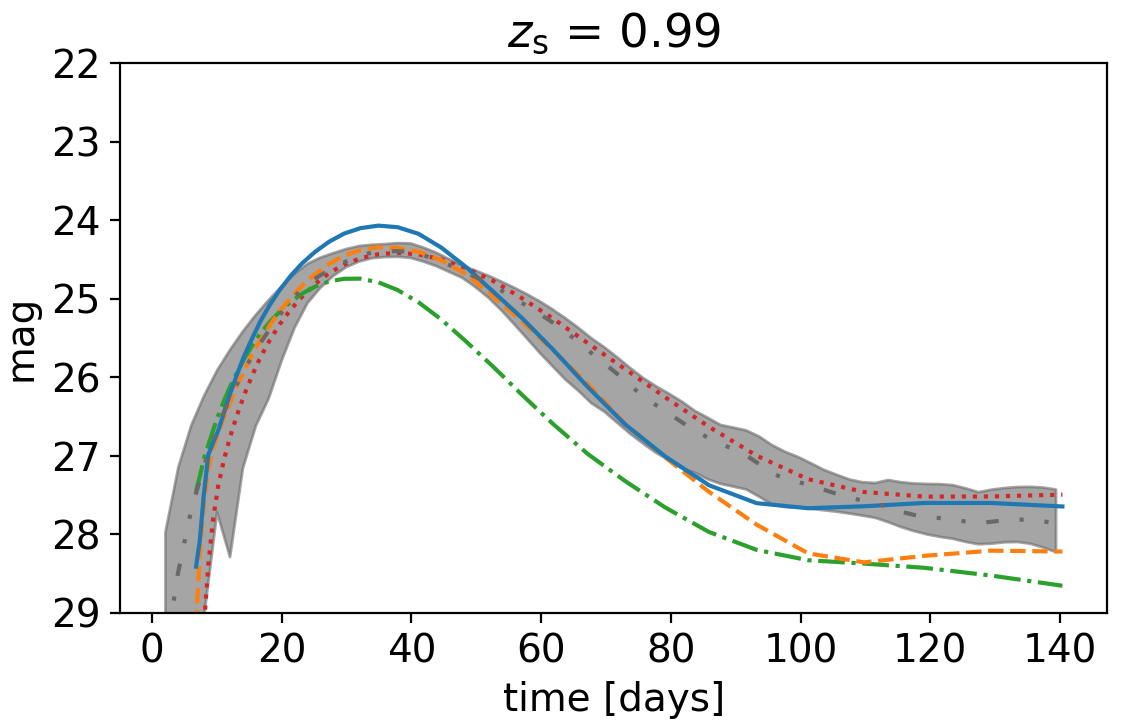}}
\caption{$i$ band light curves for typical source redshifts of LSNe Ia as indicated on the top of each panel. We compare four theoretical models (merger, N100, sub-Ch, and W7) with the empirical \texttt{SNEMO15} model.}
\label{fig:i_band_light_curves_theo_model_vs_emp}
\end{figure}

The theoretical light curves used in the training process match only partly the empirical \texttt{SNEMO15} light curves as can be seen from Fig. \ref{fig:i_band_light_curves_theo_model_vs_emp} for typical source redshifts. To increase the variety of the light curve shapes from theoretical models we multiply a random factor between 0.7 and 1.3 to the time after explosion to stretch and squeeze the light curves and add a random shift in magnitude between $-$0.4 and 0.4 to vary the brightness, given that \cite{Huber:2021iug} have shown that this helps the FCNN and the RF best to generalize better to SN models not used in the training process. This leads to absolute magnitudes from $-17.9$ to $-19.7$ in $u$ band and $-18.7$ to $-19.7$ in $g$ band. Further, our underlying light curves from theoretical and empirical models cover a range from $\sim$3.4 to 70.0 rest-frame days after explosion.

Finally, to create a training, validation, or test set, we pick randomly a time delay between 0 to 150 days for each LSN Ia system, and decide  if our LSN Ia light curves will be based on a theoretical model or the empirical model (including microlensing from a theoretical model) with equal probability. For the theoretical model, we decide with a 50:50 chance if we use the theoretical models as shown in Fig. \ref{fig:i_band_light_curves_theo_model_vs_emp} or if we apply a random stretching factor between 0.7 and 1.3 and a random shift in magnitude between $-$0.4 and 0.4. Further, we pick randomly for the first image a minimum ($(\kappa,\gamma) = (0.37,0.35),(0.27,0.24), {\rm or\ } (0.46,0.48)$) and for the second image a saddle ($(\kappa,\gamma) = (0.69,0.7), (0.9,0.9), {\rm or\ } (0.52,0.53)$), as well as a source redshift from the OM10 sample, for which we approximate the microlensing contribution following the process described in Eqs. (\ref{eq:redshift corrected flux}) to (\ref{eq:redshift corrected time}). Given that the empirical \texttt{SNEMO15} model covers rest-frame wavelengths only between $\SI{3305}{\angstrom}$ to $\SI{8586}{\angstrom}$, we cannot calculate $i$-band light curves for $\sourcez > 1.0$; therefore, if the empirical model is picked for such a high $\sourcez$, we draw the redshift again until we have $\sourcez \le 1.0$. Furthermore, we draw a random position in the moon phase and assume the detection and follow-up strategy as described previously. For this specific random setup, we create then four samples as described by Eq. (\ref{eq:sample}) for the four theoretical SN Ia models, or if the empirical model is picked we use, e.g., for the training set one of the 87 \texttt{SNEMO15-T} light curves, where we create also four samples with four different microlensing contributions from the four theoretical models. For this specific setup we calculate then the microlensed SN Ia light curves with random noise following Eq. (\ref{eq:noise realization random mag including error LSST science book}), leading to samples shown on the left hand side of Fig. \ref{fig:random_examples_training_data}. We repeat the previously described procedure by drawing new time delays until we reach the target size of a certain data set. For the training set, we use 200000 samples, and the validation and test sets contain each 20000 samples.

\section{Machine learning technique}
\label{sec:Machine learning technique used in this work}

In Sect. \ref{sec:basics of LSTM} we summarize briefly the basic concept of an LSTM network before we introduce our setup in Sect. \ref{sec:Modified Siamese Long Short-Term Memory Network}.

\subsection{Basics of Long Short-Term Memory Network}
\label{sec:basics of LSTM}

\begin{figure}[htbp]
\centering
\includegraphics[width=0.2\textwidth]{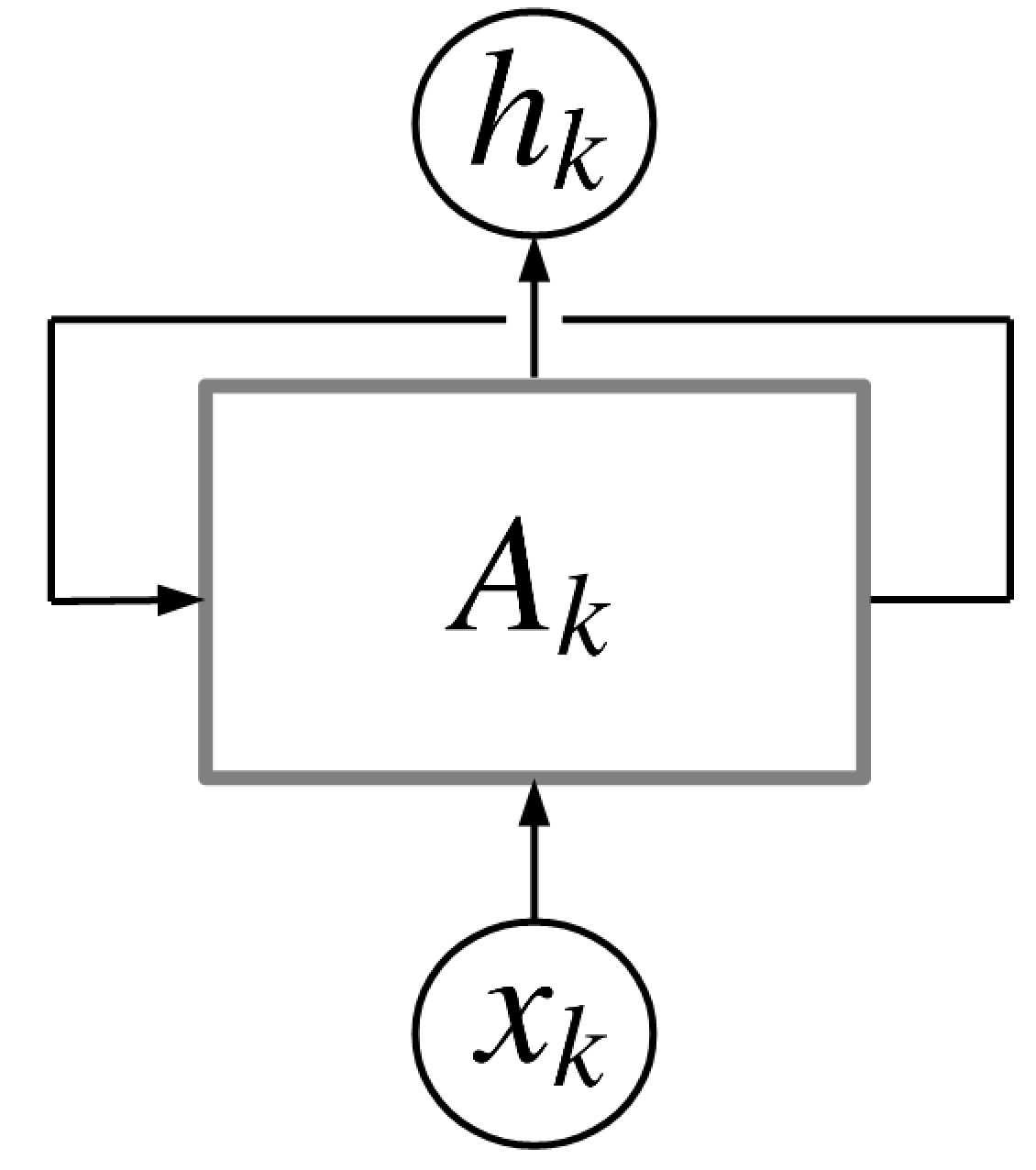}
\caption{Illustration of an RNN. The input $x_k$ is used to calculate state $A_k$ of the network at the $k$-th time step, by using also the information from the previous $k - 1$ time step. The output $h_k$ is then computed from the state $A_k$.  See Eqs.~(\ref{eq:Ak}) and (\ref{eq:hk}).}
\label{fig:RNN}
\end{figure}

The LSTM network \citep{hochreiter1997long,sak2014long,Sherstinsky_2020} is motivated by the Recurrent Neural Network (RNN), where both have the purpose to solve a time-dependent problem. For the RNN, the basic setup is illustrated in Fig. \ref{fig:RNN} for the $k$-th time step. We have an input vector $x_k$ for which we calculate the state $A_k$ by using information of the previous time step $k-1$ to further calculate the output $h_k$. 
In equations, ignoring non-linear activation functions, this can be expressed as 

\begin{equation}
\label{eq:Ak}
 A_k = w_A A_{k-1} + w_x x_k 
\end{equation}
and
\begin{equation}
\label{eq:hk}
 h_k = w_h A_{k},
\end{equation}
where $w_A$, $w_x$, and $w_h$ are the time-step independent weights which are learned in the training process via backpropagation. The main issue with RNN is that they have the tendency of exploding or vanishing gradients, which makes it very difficult to learn long-term dependencies \citep{Pascanu2013OnTD,Philipp2018GradientsE}. 

\begin{figure}[htbp]
\centering
\includegraphics[width=0.49\textwidth]{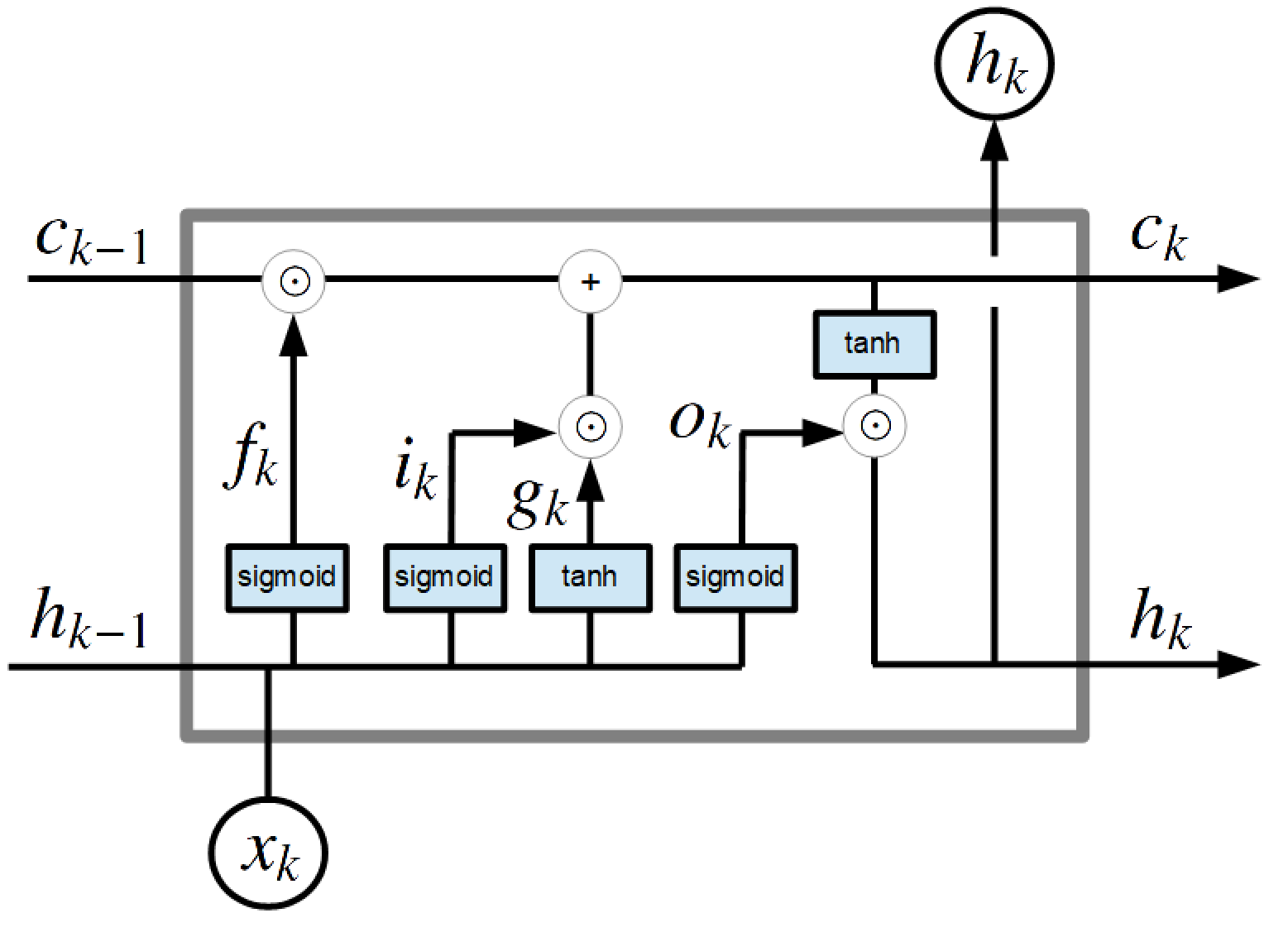}
\caption{Illustration of an LSTM. The input $x_k$ is used in combinations with the output $h_{k-1}$ and the cell state $c_{k-1}$ from the previous time step $k-1$ to produce the output $h_k$ for the $k$-th time step. The cell state $c_k$ acts as memory and transports information through the cell, where the forget gate $f_k$, input gate $i_k$, and gate $g_k$ update information in the cell state. The output gate, based on the previous output $h_{k-1}$ and input $x_{k}$ in combination with information from the cell state provides the new output $h_k$. For more information, see Eqs. (\ref{eq: first equation LSTM}) to (\ref{eq: last equation LSTM}).}
\label{fig:LSTM}
\end{figure}

To overcome that issue, LSTM network have been introduced by \cite{hochreiter1997long}, where we have in addition to the hidden state $h_k$, a cell state $c_k$, which acts as a memory. Furthermore, the state $A_k$ is replaced by a more complex unit as shown in Fig. \ref{fig:LSTM}, which contains the following main ingredients with $\odot$ denoting the Hadamard product\footnote{the Hadamard product is an element-wise product for two matrices of the same dimension}:

\begin{equation}
 h_k = o_k \odot \mathrm{tanh}(c_k),
 \label{eq: first equation LSTM}
\end{equation}
\begin{equation}
 c_k = f_k \odot c_{k-1} + i_k \odot g_k,
\end{equation}
\begin{equation}
 f_k = \mathrm{sigmoid}(w_{xf} x_k + w_{hf} h_{k-1} + b_f),
\end{equation}
\begin{equation}
 g_k = \mathrm{tanh}(w_{xg} x_k + w_{hg} h_{k-1} + b_g),
\end{equation}
\begin{equation}
i_k = \mathrm{sigmoid}(w_{xi} x_k + w_{hi} h_{k-1} + b_i),
\end{equation}
\begin{equation}
 o_k = \mathrm{sigmoid}(w_{xo} x_k + w_{ho} h_{k-1} + b_o),
\label{eq: last equation LSTM}
\end{equation}
where the sigmoid and tanh activation functions act on each element of the vector. The input vector at time step $k$ is $x_k$ with size $N_\mathrm{input}$, which is, in our case, three, containing the time, magnitude, and uncertainty as listed in Eq. (\ref{eq:data structure time, mag, unc}). The hidden state $h_{k}$ is a vector with size $N_\mathrm{hidden}$. The biases $b_i$, $b_f$, $b_g$, and $b_o$ are vectors with size $N_\mathrm{hidden}$, the weights $w_{xf}$, $w_{xg}$, $w_{xi}$, and $w_{xo}$ are $N_\mathrm{hidden}  \times N_\mathrm{input}$ matrices, and the weights $w_{hf}$, $w_{hg}$, $w_{hi}$, and $w_{ho}$ are $N_\mathrm{hidden}  \times N_\mathrm{hidden}$ matrices. The biases and weights are not dependent on the time step and are learned through backpropagation in the training process.
The hidden state $h_k$ is the prediction of the LSTM network at time step $k$ and serves as input to calculate the hidden state $h_{k+1}$ corresponding to the time step $k+1$. The cell state $c_k$ acts as memory and transports information through the unit. The forget gate $f_k$ decides on when and which parts to erase from the cell state. The gate $g_k$ removes or adds information to the cell state coming from the input gate $i_k$. The output gate $o_k$ is responsible for the output / hidden state $h_k$ of the LSTM network. Outputs are produced at every time step, but for our case, we will only use the last output at time-step $N_\mathrm{sl}$, after the full light curve with the sequence length of $N_\mathrm{sl}$ was propagated through the network. This corresponds to a many-to-one scenario where we have $N_\mathrm{sl}$ input data points as in Eq. (\ref{eq:data structure time, mag, unc}) for which we predict one time delay $\Delta t$ and the uncertainty of the time delay $\sigma_{\Delta t}$.

\subsection{Modified Siamese Long Short-Term Memory Network}
\label{sec:Modified Siamese Long Short-Term Memory Network}

For our approach, we use a Siamese LSTM network \citep{doi:10.1142/S0218001493000339}. The Siamese neural networks are in general a class of neural network architectures that contain two or more identical sub-networks, which therefore have the same weights and parameters.
A Siamese LSTM is also used in the Manhattan LSTM (MaLSTM) network \citep{Mueller_Thyagarajan_2016} developed to find semantic similarities between sentences. In this network two sentences with different (or the same) number of words are feed through the same LSTM network to predict an output for each sentence. The two outputs are then used to calculate a Manhattan distance which states the similarity between the two input sentences.

Our approach is similar to the MaLSTM, but instead of sentences, we have different images of a LSN Ia, and instead of a single word, we have the time of the observation, the magnitude, and uncertainty of the measurement as shown in Eq. (\ref{eq:data structure time, mag, unc}). Our approach produces similar to the MaLSTM two outputs for the two images, but instead of using a Manhattan distance, we use a FCNN with two hidden layers, given that we want to determine a time delay instead of a similarity score. 

\begin{figure*}[htbp]
\centering
\includegraphics[width=0.9\textwidth]{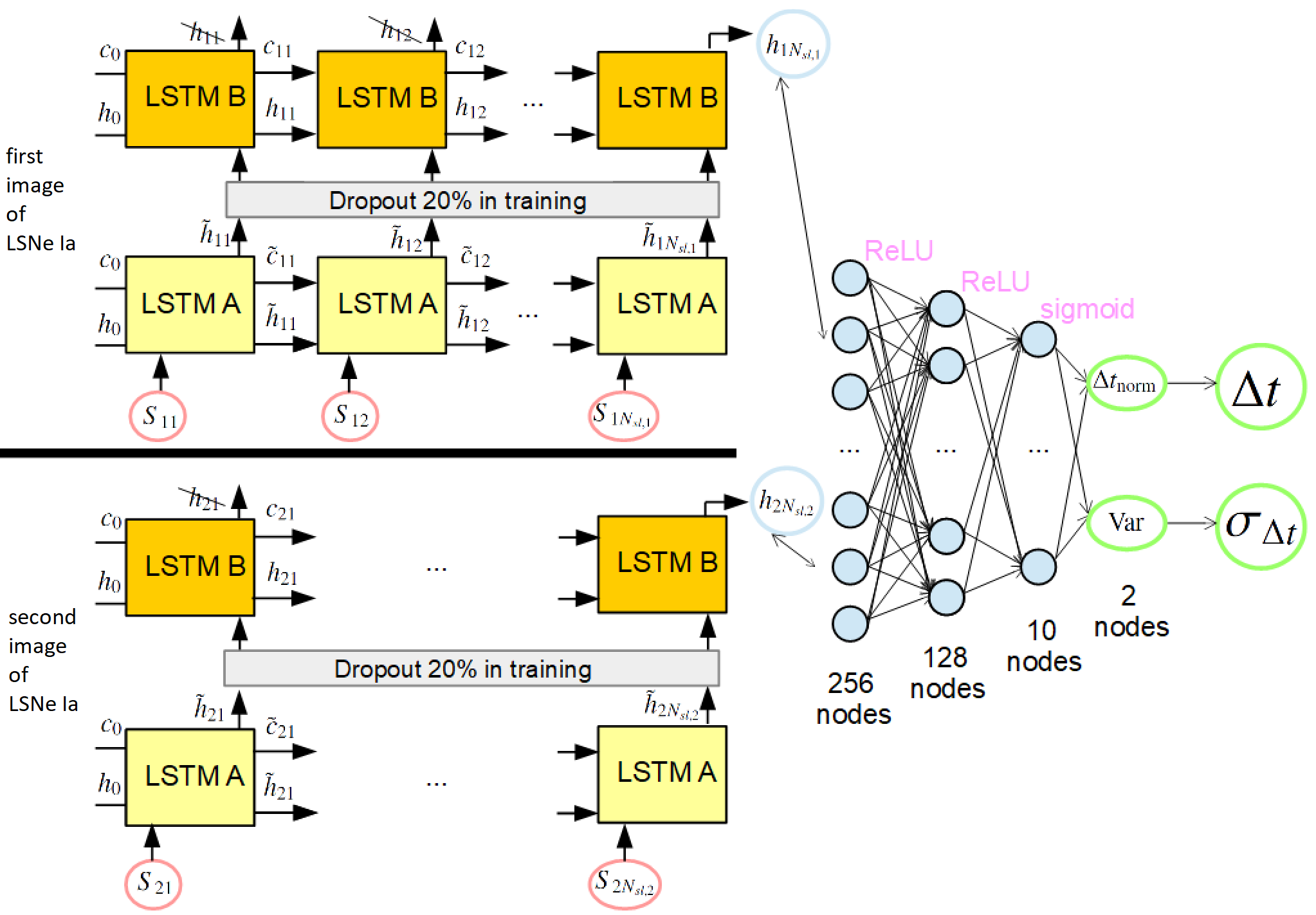}
\caption{Illustration of our \slstm\ composed of a Siamese LSTM network and a FCNN with two hidden layers. Details about the LSTM cells are shown in Fig. \ref{fig:LSTM}. Both sub-LSTM networks for image one (top, input denoted by $S_{1k}$) and image two (bottom, input denoted by $S_{2k}$) are the same LSTM network (i.e., with the same weights in the neural network) composed of 2 layers (LSTM A and LSTM B) where during the training process, a dropout rate of 20\% is used. We use only the last outputs $h_{1N_\mathrm{sl,1}}$ and $h_{2N_\mathrm{sl,2}}$ from the second layer. These outputs are concatenated to be the input for the FCNN with two hidden layers using ReLU activation functions and a sigmoid activation for the final output. Given that the data is normalized the output of the network, $t_\mathrm{norm}$ and Var, needs to be converted to get the time delay $\Delta t$ (see Eq. (\ref{eq: final output of SLSTM}) and the uncertainty of the time delay $\sigma_{\Delta t}$ (see Sec. \ref{sec: Results}). The initial cell state $c_0$ and hidden state $h_0$ have the size $N_\mathrm{hidden}$ with all elements being equal to zero.}
\label{fig:LSTM_Siamese_FCNN}
\end{figure*}

Our full \slstm\ is shown in Fig. \ref{fig:LSTM_Siamese_FCNN}, where the calculations of the LSTM network are listed in detail in Eq. (\ref{eq: first equation LSTM}) to (\ref{eq: last equation LSTM}), where we added a second index $j$ to the the time step $k$, leading to $h_{jk}$, $c_{jk}$, $f_{jk}$, $g_{jk}$, $i_{jk}$, and $o_{jk}$ in order to distinguish between image one ($j=1$) and image two ($j=2$) of our lensed SN system. Our network takes as input the data $S_{1k}$ of image one and $S_{2k}$ of image two, and predicts as an output, the time delay $\Delta t$ and the uncertainty of the time delay $\sigma_{\Delta t}$. The sub-LSTM networks for both images are identical and composed of the two layers LSTM A and LSTM B. The second layer of the LSTM network takes as input, the output from the first layer $\tilde{h}_{jk}$, where a dropout rate $p_\mathrm{dropout} = 20\%$ is applied, meaning that during the training process randomly 20\% of the input data for the second layer ($\tilde{h}_{jk}$) are set to zero and the other values are rescaled by 1/(1-$p_\mathrm{dropout}$). The rescaling is necessary given that the dropout is just applied in the training process but not during the validation and test phase, which would lead to systematically larger outputs for the validation and test set if the re-scaling in the training process would not be applied. The dropout helps the network to avoid focusing on very specific nodes during training in order to generalize better to data not used in the training process. The last two outputs from the second layer $h_{1N_\mathrm{sl,1}}$ for image one and $h_{2N_\mathrm{sl,2}}$ for image two are then concatenated and used as input for the FCNN.

For given $j$ and $k$, the inputs $S_{jk}$ are vectors with size of 3 (see Eq.~(\ref{eq:data structure time, mag, unc})) whereas the outputs from both layers $\tilde{h}_{jk}$ and $h_{jk}$ are vectors with size $N_\mathrm{hidden} = 128$. Therefore, the input layer of the FCNN has 256 nodes, which we summarize in a vector $d$: 
\begin{equation}
d := (h_{1N_\mathrm{sl,1}},h_{2N_\mathrm{sl,2}}),
\end{equation}
with dimension 256. The two hidden layers of the FCNN have 128 and 10 nodes, where at each hidden layer, a non-linearity is applied by using a rectified linear units (ReLU) activation function \citep[e.g.,][]{Schmidt:1998,Maas:2013}, which is the identity function for positive values and zero for all negative values. In the last layer where final outputs are produced, a sigmoid activation is used which predicts values between 0 and 1 and therefore covers the range of potential normalized time delays.

The first hidden layer is represented by the vector $\eta_{1}$ with size 128 and the second hidden layer by the vector $\eta_{2}$ with size 10:

\begin{equation}
\eta_{1} =  \mathrm{ReLU}\bigr(\omega_{1} \, d + \beta_{1}\bigl),
\end{equation}
\begin{equation}
\eta_{2} =  \mathrm{ReLU}\bigr(\omega_{2} \, \eta_{1} + \beta_{2}\bigl).
\end{equation}
The outputs of the network are then two scalar values, namely the normalized time delay $\Delta t_\mathrm{norm}$ and the normalized variance ``$\mathrm{Var}$'':
\begin{equation}
(\Delta t_\mathrm{norm}, \mathrm{Var}) = \mathrm{sigmoid}\bigr(\omega_{3} \, \eta_{2} + \beta_{3}\bigl),
\label{eq: final output of SLSTM}
\end{equation}
where, $\omega_{1}$ (matrix of size $128 \times 256$), $\omega_{2}$ (matrix of size $10 \times 128$), and $\omega_{3}$ (matrix of size $2 \times 10$) are the weights, and $\beta_1$ (vector of size 128), $\beta_{2}$ (vector of size 10), and  $\beta_{3}$ (vector of size 2) are the biases. Weights and biases of the network are learned in the training process. Further, the outputs from Eq. (\ref{eq: final output of SLSTM}) need to be translated to a physically meaningful $\Delta t$ and $\sigma_{\Delta_t}$. For the time delay, this is straight forward by just removing our normalization factor:
\begin{equation}
\Delta t = \Delta t_\mathrm{norm} \cdot 150 \, \mathrm{days},
\label{eq: time delay remove normalization}
\end{equation}
but for $\sigma_{\Delta_t}$ this is not as trivial and we will explain our approach in Sec. \ref{sec: Results}. 

For the implementation of the \slstm, we use the machine learning library \texttt{PyTorch} \citep{NEURIPS2019_9015}, where we train for a certain number of epochs until a generalization gap becomes visible (more details in Sect. \ref{sec: Results}). In each epoch, we subdivide our training set randomly into mini batches with size $N_\mathrm{batch} = 128$. Given that image one and image two of the 128 samples (labeled as $l$) of a mini batch have a different sequence length $N_{\mathrm{sl,1},l}$ and $N_{\mathrm{sl,2},l}$, we pad them with zeros at the end until all mini-batch samples of image one have a shape of (max($N_{\mathrm{sl,1},l}$), 3) and all samples of image two have a shape of (max($N_{\mathrm{sl,2},l}$), 3), where the 3 is coming from the three inputs from Eq. (\ref{eq:data structure time, mag, unc}). The padding is required such that the \texttt{PyTorch} class \texttt{torch.nn.LSTM} can handle variable sequence lengths within a mini batch\footnote{to implement this, the \texttt{PyTorch} functions \texttt{torch.nn.utils.rnn.pad\_sequence}, \texttt{torch.nn.utils.rnn.pack\_padded\_sequence}, and \texttt{torch.nn.utils.rnn.pad\_packed\_sequence} are used}.  As input for the FCNN, we use for each sample $l$ the last \textit{unpadded} output at time step $N_{\mathrm{sl,1},l}$ for image one and $N_{\mathrm{sl,2},l}$ for image two, which are then combined to predict $\Delta t$ and $\sigma_{\Delta t}$. For each sample of the mini batch, we can then compute the performance of the \slstm\ by using the Gaussian negative log likelihood \citep{374138} as the loss function:

\begin{equation}
\mathrm{loss} = 0.5 \bigl(\mathrm{log}(\mathrm{max}(\mathrm{Var}, \epsilon)) + \frac{(\Delta t_\mathrm{norm}-\Delta t_\mathrm{norm,true})^2}{\mathrm{max}(\mathrm{Var, \epsilon)}}\bigr),
\label{eq:loss_function}
\end{equation}
where $\epsilon = 10^{-6}$, is used for stability. The true normalized time delay of the training sample is $\Delta t_\mathrm{norm,true}$ and $\Delta t_\mathrm{norm}$ is the time delay predicted from the network. Furthermore, we use the second ``raw'' output of the \slstm\ as the variance ``Var'' as in Equation \ref{eq: final output of SLSTM}. The advantage of the Gaussian negative log likelihood loss in comparison to the mean squared error loss used in \cite{Huber:2021iug} is that we not only optimize such that the predicted time delay of the network is close to the true time delay, but we also get an approximation of the uncertainty for each sample. The loss is then used for the backpropagation to update the weights and biases of the \slstm. For this we use the Adaptive Moment Estimation (Adam) optimizer \citep{Kingma:2014} with a learning rate $\alpha = 0.000005$.

The network with our specific choices/values of $N_\mathrm{hidden}$, $N_\mathrm{batch}$, $\alpha$, number of LSTM layers, dropout rate, FCNN with two hidden layers and so on is of course somewhat 
arbitrary, although the setup of the FCNN was motivated by \cite{Huber:2021iug}. However, we also investigated other setups for a smaller training sample, e.g., $N_\mathrm{hidden} = 64, 256, 512$, $ N_\mathrm{batch} = 64, 256, 512$, just one LSTM layer without dropout and so on, and we found that the performance of these networks were typically very similar or $\sim$10 percent worse than our chosen setup. Therefore, given that the training of a single setup of such an LSTM network takes several weeks on a GPU node (NVIDIA A10) and that we found only minor performance differences for a few tested configurations of the \slstm, we skipped a huge grid search for all potential hyperparameters, which is 
also justified by the performance of our \slstm\ already achieving our target precision and accuracy.

\section{Results}
\label{sec: Results}

In Sect. \ref{sec:Results Training process} we describe how the \slstm\ is trained, in Sect. \ref{sec:Results performance} we evaluate the performance by applying it to our test sets and in Sect. \ref{sec:Results Uncertainty estimate} we describe the uncertainty prediction of the network.

\subsection{Training process}
\label{sec:Results Training process}

To evaluate the performance of our \slstm\ on a given data set (e.g., training set, validation set, test set), we calculate for each sample as in Eq. (\ref{eq:sample}), now labeled as $l$, the time-delay deviation via,

\begin{equation}
 \tau_{l} = \Delta t_{\mathrm{true},l} - \Delta t_{\mathrm{pred},l},
 \label{eq:time_delay_deviation}
\end{equation}
comparing the true time delay $\Delta t_{\mathrm{true},l}$ to the predicted one from the network $\Delta t_{\mathrm{pred},l}$. Given that our data is normalized in time for the input/output of our \slstm, we need to undo the normalization as in Eq. \ref{eq: time delay remove normalization} to get the time delays in days.
We can then estimate the bias (also referred to as accuracy) by using the median $\tilde{\tau}$ and the precision by using the  84th percentile $\tau_{p,84}$ and the 16th percentile $\tau_{p,16}$ from the whole sample (200000 for the training data and 20000 for each of the other data sets). The results are then summarized in the following form:
\begin{equation}
\tilde{\tau}^{\tau_{p,84}-\tilde{\tau}}_{{\tau_{p,16}} - \tilde{\tau}}.
\label{eq:bias_and_precision}
\end{equation}

We train our \slstm\ as described in Sect. \ref{sec:Modified Siamese Long Short-Term Memory Network} with the data set as described in Sect. \ref{sec:Data set for a LSTM Network} using light curves from theoretical and the empirical \texttt{SNEMO15} models as listed in Table \ref{tab:train validation test set for theo plus empirical in training process}.

\begin{figure}[htbp]
\centering
\includegraphics[width=0.49\textwidth]{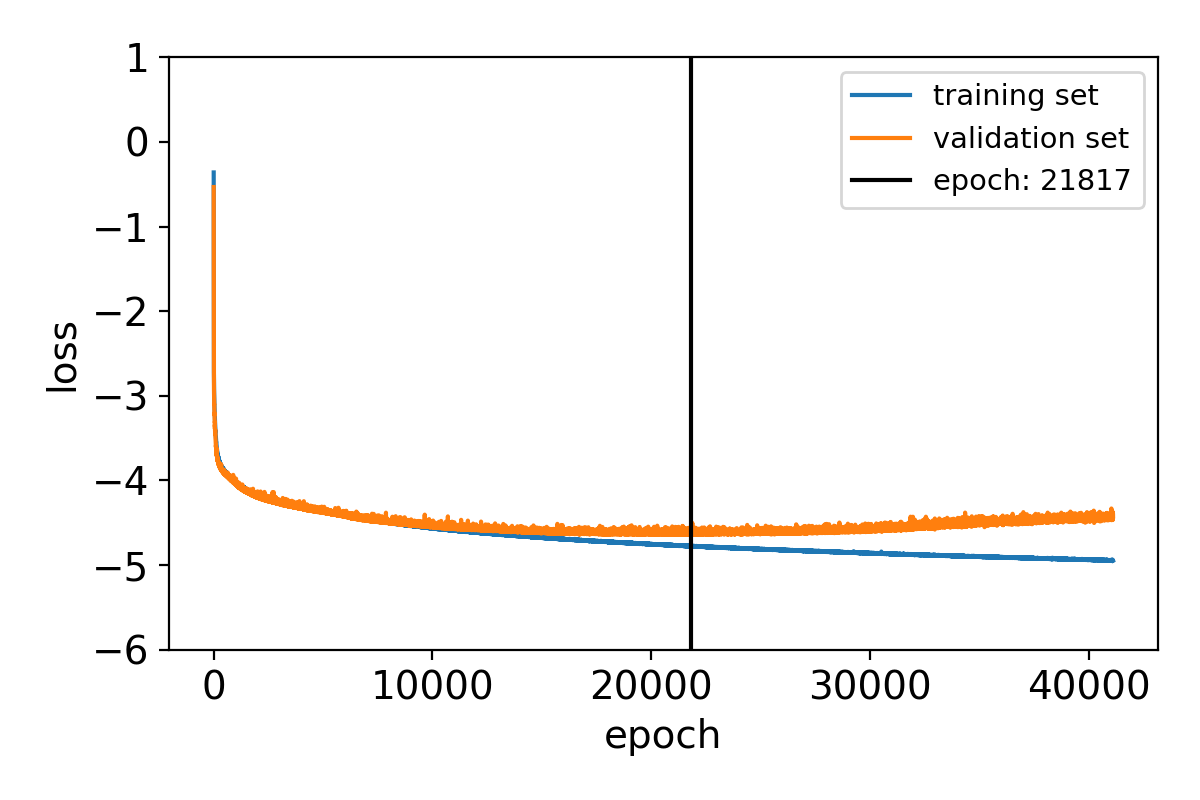}
\caption{Training and validation loss of the \slstm\ as a function of the training epoch. The vertical line marks the state at which we use the \slstm\ where the validation loss is the lowest with the additional requirement that the bias is lower than 0.05 day. The plot shows that with more training epochs the network improves its performance while around epoch 22000 the validation loss starts to increase again while the training loss still decreases because the \slstm\ starts to overfit, meaning that, e.g., it fits to specific noise patterns present in the training set.}
\label{fig:loss_training_vs_validation}
\end{figure}

Our training process lasts for $\sim$40000 epochs and we use the \slstm\ at the epoch where the validation set has the minimum loss value calculated from Eq. (\ref{eq:loss_function}) with the additional requirement that the bias as defined in Eq. (\ref{eq:bias_and_precision}) at this epoch is lower than 0.05 day, to ensure that at least for the validation set, a time delay longer than 5 days would be enough to achieve a bias lower than 1\%. The loss curve for the training and validation set as a function of the training epoch is shown in Fig.~\ref{fig:loss_training_vs_validation}, where we see the generalization gap 
beyond epoch $\sim$15000, and we reach the lowest validation loss with low bias at epoch 21817, which is the status of the \slstm\ as used further in this work.

\subsection{Performance}
\label{sec:Results performance}
\begin{figure}[htbp]
\centering
\includegraphics[width=0.49\textwidth]{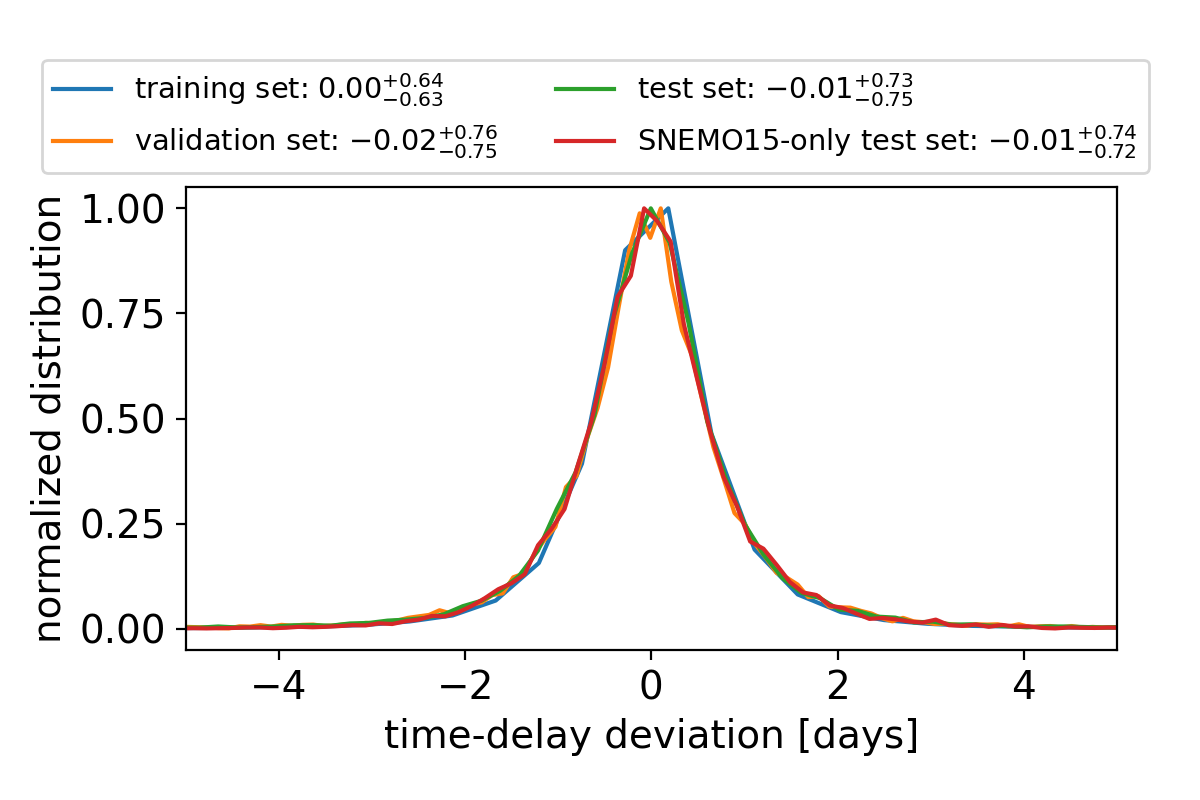}
\caption{Evaluation of the \slstm\ on the training set, validation set, test set, and \texttt{SNEMO15}-only test set. With a $1\sigma$ spread of $\sim$0.7 days and almost no bias even on the two test sets, the \slstm\ is very promising for future measurements of time delays from real LSNe Ia.}
\label{fig:final_result}
\end{figure}

 For our \slstm\ we calculate the time-delay deviation for the training set, validation set, test set, and \texttt{SNEMO15}-only test set which is shown in Fig. \ref{fig:final_result}. We see that our network performs very well with almost zero bias and a precision on the order of 0.7 days, which can be also achieved on light curve shapes never used in the training process, as represented by the \texttt{SNEMO15}-only test set.
 
 Our \slstm\ is constructed such that the image which arrives first needs to be the first input and the second arriving image is required as the second input. For short time delays ($\Delta t \lesssim 10\, \mathrm{days}$) it might not always be possible to distinguish between the first and second image. However, in practice this is not a problem as the \slstm\ ouputs typically a value very close to zero if the input of the two images has the wrong order. Therefore, when we test our network, we try as input both potential orders of the two images and always take the higher time delay as the predicted one. This is already done for the test set and SNEMO15-only test set in Fig. \ref{fig:final_result}. The presented result would be exactly the same if we input the images always in the right order, and therefore, we see that this limitation of our \slstm\ is not a problem in practice.

\subsection{Uncertainty estimate}
\label{sec:Results Uncertainty estimate}

\begin{figure}[htbp]
\centering
\includegraphics[width=0.49\textwidth]{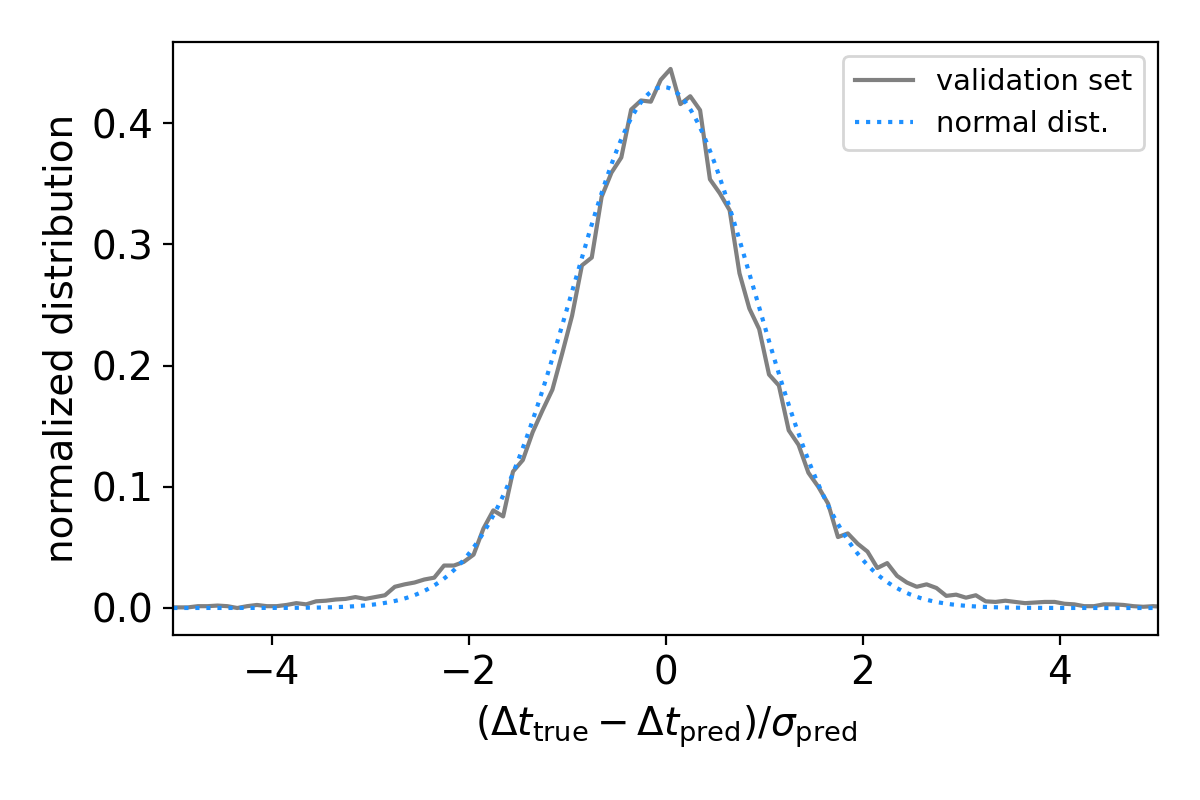}
\caption{
Application of the \slstm\ to the validation set. The distribution shows the time-delay deviation of each sample over the predicted uncertainty of the network. In blue we see a fit with a normal distribution.}
\label{fig:uncertainty lstm on validation set}
\end{figure}

\begin{figure}[htbp]
\centering
\includegraphics[width=0.49\textwidth]{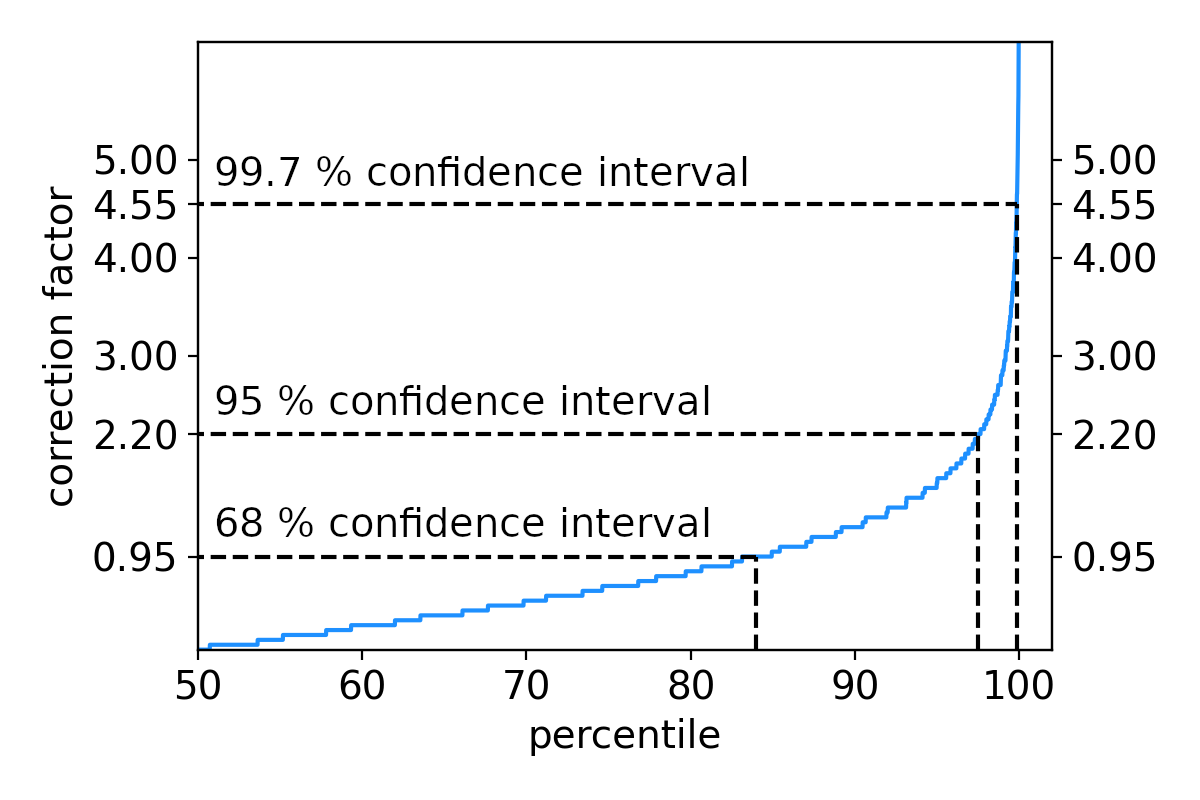}
\caption{Based on Fig. \ref{fig:uncertainty lstm on validation set} we can define confidence intervals and find based on the fit function $f$ a correction factor that needs to be applied to the predicted uncertainty of the network $\sigma_\mathrm{pred}$.}
\label{fig:uncertainty correction factor}
\end{figure}

In order to relate the raw Var output from our \slstm\ also listed in Eq. (\ref{eq:loss_function}) to a meaningful $\sigma_{\Delta t}$, we use the validation set as shown in Fig. \ref{fig:final_result}. To convert the Var output, we first calculate the rescaled uncertainty prediction from the network 

\begin{equation}
\sigma_\mathrm{pred} = \sqrt{\mathrm{Var}} \cdot 150 \, \mathrm{days},
\label{eq: rescaled sigma values output of LSTM}
\end{equation}
where the factor 150 comes from the normalization of our time scale as described in Sect. \ref{sec:Data set for a LSTM Network}. In a next step, we apply the \slstm\ to the full validation set and plot for each sample the time delay deviation (as defined in Eq. (\ref{eq:time_delay_deviation})) over the rescaled predicted uncertainty of the network $\sigma_\mathrm{pred}$ as shown in Fig. \ref{fig:uncertainty lstm on validation set}, where we fit a normal function to it. Although the normal function fits the distribution overall well, the outer regions, where the predicted time delay $\Delta t_\mathrm{pred}$ is more than 2$\sigma_\mathrm{pred}$ away from the true time delay $\Delta t_\mathrm{true}$, are not represented very well. Therefore, we look at different confidence intervals of the validation set which yields a correction factor as shown in Figure \ref{fig:uncertainty correction factor}. The final uncertainty estimate for the 64th, 95th and 99.7th confidence interval of a single sample can then be calculated via 

\begin{equation}
\sigma_\mathrm{84th-16th} \approx 0.95 \sigma_\mathrm{pred},
\label{eq:uncertainty one sigma confidence interval}
\end{equation}

\begin{equation}
\sigma_\mathrm{97.5th-2.5th} \approx 2.20 \sigma_\mathrm{pred},
\end{equation}
and
\begin{equation}
\sigma_\mathrm{99.85th-0.15th} \approx 4.55 \sigma_\mathrm{pred}.
\end{equation}

\section{Specific applications of the \slstm\, and comparison to other methods}
\label{sec:specific applications of LSTM network}

\begin{figure}[htbp]
\centering
\includegraphics[width=0.49\textwidth]{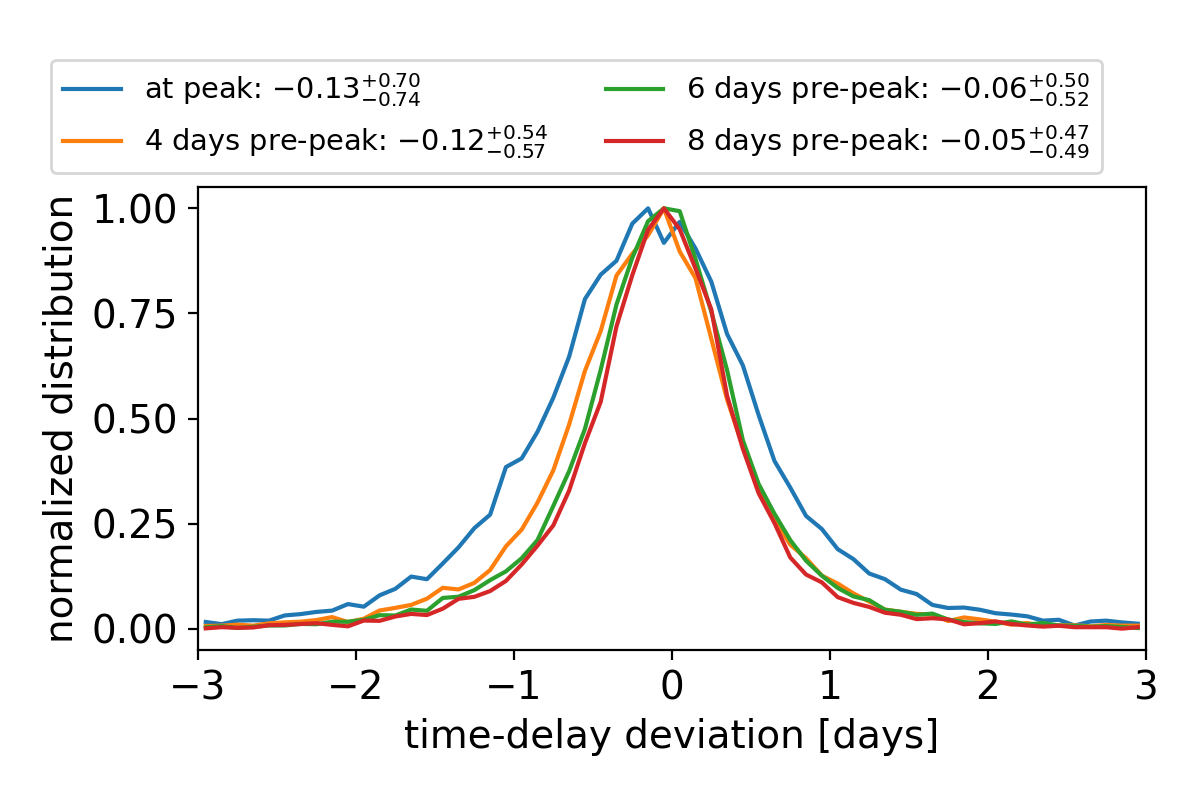}
\caption{Evaluation on a specific LSN Ia system at $\sourcez = 0.76$ with $(\kappa_1,\gamma_1,s_1)=(0.25, 0.27, 0.6)$  for image one and $(\kappa_2,\gamma_2,s_2)=(0.83, 0.81, 0.6)$ for image two as in \cite{Huber:2021iug} for at peak detection and three cases where the first SN image is detected before peak (in observer-frame days). The light curves are simulated in the same way as those in the \texttt{SNEMO15}-only test set. As expected the precision drops with less available data points pre-peak, but the uncertainty of $\sim$0.7 day for at-peak detection is still good. The accuracy is worse than in Figure \ref{fig:final_result}, which shows that the spread presented there is not only due to various noise and microlensing configurations but also comes from different types of LSNe Ia, although the bias cancels out if a large variety of systems is investigated.} 
\label{fig:comparison_holismokes_7_detection_before_peak}
\end{figure}

In this Section, we apply the \slstm\ to more specific examples to compare the performance to the results from \cite{Huber:2021iug}. In particular, we create a data set as described in Sect.~\ref{sec: Data set for machine learning}, but we fix $\sourcez, \kappa, \gamma,$ and $s$ to very specific values as used by \cite{Huber:2021iug}, where a system was picked from the OM10 catalog with a source redshift $\sourcez = 0.76$ that is closest to the median source redshift $\sourcez = 0.77$ of the OM10 catalog. The specific system is a double LSN Ia with $(\kappa_1,\gamma_1,s_1)=(0.25, 0.27, 0.6)$ for the first SN image and $(\kappa_2,\gamma_2,s_2)=(0.83, 0.81, 0.6)$ for the second SN image. For this system \cite{Huber:2021iug} investigated different scenarios of at peak and before peak detection. Figure \ref{fig:comparison_holismokes_7_detection_before_peak} shows our \slstm\ on a \texttt{SNEMO15}-only test set, based on $(\kappa_1,\gamma_1,s_1)=(0.25, 0.27, 0.6)$ and $(\kappa_2,\gamma_2,s_2)=(0.83, 0.81, 0.6)$ for four different times of detection.

We first see that even though the specific
$\kappa$ and $\gamma$ values are not part of our training data (Table \ref{tab:random parameters for LSTM training}), the network predicts very accurate and precise time delays. Further, we see that the precision drops with fewer data points before peak, although at-peak detection still yields an uncertainty of $\sim$0.7 day, which is still a precise time-delay measurement. Furthermore, we find that the accuracy deviates from zero, showing that the spread detected in Fig.~\ref{fig:final_result} not only comes from random noise and microlensing positions -- for very specific systems, the \slstm\ slightly over or underestimates the predicted time delay, which averages out if large samples of LSNe Ia are investigated.

\begin{figure}[htbp]
\centering
\includegraphics[width=0.49\textwidth]{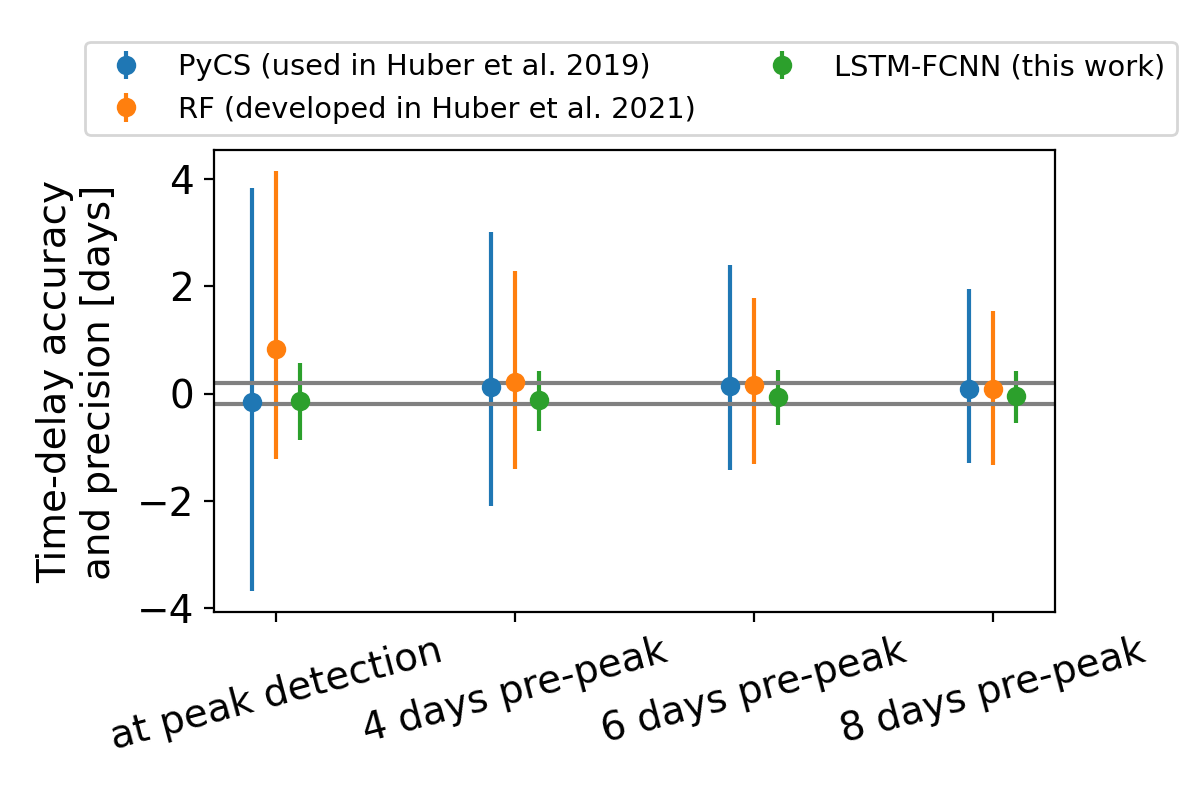}
\caption{The \slstm\ in comparison to the RF developed in \cite{Huber:2021iug} and \texttt{PyCS} used in \cite{Huber:2019ljb}. The exact values are listed in Table \ref{tab:LSTM-FCNN vs RF vs PyCS detection before peak}. In terms of precision, the \slstm\ outperforms the other two approaches by roughly a factor of three.}
\label{fig:comparison_holismokes_7_detection_before_peak2}
\end{figure}

In Figure \ref{fig:comparison_holismokes_7_detection_before_peak2} (exact values are listed in Table \ref{tab:LSTM-FCNN vs RF vs PyCS detection before peak}), we compare the results from Fig.~\ref{fig:comparison_holismokes_7_detection_before_peak} to similar runs for the RF and \texttt{PyCS} taken from \cite{Huber:2021iug}. We find that the accuracy of the \slstm\ is comparable to \texttt{PyCS} and better than the RF, especially if fewer data points are available before peak. However, if we look at the precision, we see that the \slstm\ outperforms the RF and \texttt{PyCS} by far (roughly a factor of three). \texttt{PyCS} provides the worst precision, which is not surprising, given that splines are used and therefore any knowledge about SN Ia light curves is ignored. Although the RF outperforms \texttt{PyCS}, it is still significantly worse than the \slstm. One reason for the worse performance of the RF is that the RF was only trained on theoretical models, where the evaluation was done on the empirical \texttt{SNEMO15} model. Although the RF did generalize well to the light curve shapes not used in the training process in terms of bias, the precision dropped by $\sim$0.5 days in comparison to a test set based on theoretical light curves as used in the training process \citep{Huber:2021iug}. Even if we would improve the precision by $\sim$0.5 days by including $\texttt{SNEMO15}$ light curves in the training process of the RF, the \slstm\ would still outperform the RF by almost a factor two in terms of precision. The second reason for the much better performance seems to be that the LSTM structure is better suited to the applied problem given that it was build to solve time-dependent problems.

\begin{table}[htbp]
\caption{Time-delay deviations of the \slstm\ on \texttt{SNEMO15}-only test sets in comparison to the RF and \texttt{PyCS} \citep[results are taken from HOLISMOKES VII for \texttt{SNEMO15} data sets][]{Huber:2021iug} for a LSN Ia with ($\kappa_1,\gamma_1,s_1)=(0.25, 0.27, 0.6)$  for image one and $(\kappa_2,\gamma_2,s_2)=(0.83, 0.81, 0.6)$ for image two for $\sourcez = 0.76$ and various detection times (in observer days) before the first SN image peaks in brightness.}
\begin{tabular}{l|ccccc}
& \slstm & RF & \texttt{PyCS} \\[0.07cm] 
\hline
\\
at peak detection & $-0.13^{+0.70}_{-0.74}$ & $0.83^{+3.33}_{-2.05}$ & $-0.16^{+3.99}_{-3.52}$ \\[0.2cm]
4 days pre-peak & $-0.12^{+0.54}_{-0.57}$ & $0.22^{+2.07}_{-1.62}$ & $0.12^{+2.89}_{-2.22}$ \\[0.2cm]
6 days pre-peak & $-0.06^{+0.50}_{-0.52}$ & $0.17^{+1.61}_{-1.48}$ & $0.15^{+2.25}_{-1.57}$ \\[0.2cm]
8 days pre-peak & $-0.05^{+0.47}_{-0.49}$ & $0.09^{+1.46}_{-1.42}$ & $0.09^{+1.87}_{-1.39}$ \\[0.2cm]
\end{tabular}
\centering
\label{tab:LSTM-FCNN vs RF vs PyCS detection before peak}
\end{table}

\section{Summary}
\label{sec:Summary}

In this work we developed a \slstm\ to measure time delays of LSNe Ia which we expect to detect with LSST and for which we plan to trigger follow-up observations. With $i$-band only to a mean single-epoch 5$\sigma$ depth of 24.5, we can achieve over a broad sample of LSNe Ia a bias free time-delay measurement with a precision of $\sim$0.7 days. Our \slstm\ is general and can be applied to a variety of LSNe Ia expected from LSST. In comparison to the RF developed in \cite{Huber:2021iug} or \texttt{PyCS} used in \cite{Huber:2019ljb} we see a significant improvement in precision of roughly three.

Because of the much better performance of the \slstm\ in comparison to \texttt{PyCS}, we can expect more LSNe Ia with well measured time delay (accuracy better than 1\% and precision better than 5\%) than that predicted by \cite{Huber:2019ljb}. For the 10-year LSST survey with a baseline observing strategy, \cite{Huber:2019ljb} predict about 28 LSNe Ia with well measured time-delay (follow-up with two day cadence and LSST-like plus one $5\sigma$ depth), out of 73 LSNe Ia systems in total. Since $\sim$30\% are dropped due to bad accuracy from the total number of LSNe Ia \citep{Huber:2019ljb},  we can expect that any new time-delay measurement method with significantly improved precision would increase the number of LSNe Ia with well measured time delays by up to a factor of $\sim$1.8 ($=73 \cdot 0.7 / 28$) in comparison to \texttt{PyCS}, which will be partly achieved by the \slstm\ given that it is roughly three to four times more precise than \texttt{PyCS}.

Although we saw that our \slstm\ performs much better than the RF on a large sample of LSNe Ia for which the RF would require a training process for each observation separately, our developed network cannot be blindly applied to any LSNe Ia. If follow-up strategies deviate from our assumptions (e.g., detection, filter, cadence) in Table \ref{tab:random parameters for LSTM training}, a new \slstm\ needs to be trained. Furthermore, we see from Table \ref{tab:random parameters for LSTM training} that we assumed only positive time delays, meaning that image one should be correctly identified as the first appearing image and image two as the second one. In practice this is not a problem as always both orders can be tested and the higher time delay should be picked as the correct one given that inputs in the wrong order will provide a time delay very close to zero. 

The \slstm\ is a very promising approach to measure time delays in LSNe. However, so far the method as presented in this work is only applicable for SNe Ia and it will be interesting in the future to develop methods based on the \slstm\ which can be applied to any kind of LSNe. 
\FloatBarrier

\begin{acknowledgements}
We thank Laura Leal-Taix\'e and Tim Meinhardt for useful discussion.
SH and SHS thank the Max Planck Society for support through the Max
Planck Research Group and Max Planck Fellowship for SHS. This project has received funding from
the European Research Council (ERC) under the European Union’s Horizon
2020 research and innovation programme (grant agreement No
771776). 
This research is supported in part by the Excellence Cluster ORIGINS which is funded by the Deutsche Forschungsgemeinschaft (DFG, German Research Foundation) under Germany’s Excellence Strategy -- EXC-2094 -- 390783311.

\end{acknowledgements}

\bibliographystyle{aa}
\bibliography{Time_delay_measurement}

\FloatBarrier

\appendix

\section{Photometric uncertainties}
\label{sec:Appendix LSST uncertainty}

The photometric uncertainty $\sigma_\mathrm{X}(t)$ used in Eq. (\ref{eq:noise realization random mag including error LSST science book}) is defined as:
\begin{equation}
\sigma_\mathrm{X}^2(t) = \sigma_\mathrm{sys}^2+\sigma_\mathrm{rand}^2(t)$,
where $\sigma_\mathrm{sys}=0.005 \, \mathrm{mag}
\label{eq:noise calculation appendix formula 1}
\end{equation}
and 
\begin{equation}
\sigma_\mathrm{rand}^2(t)= \bigl[(0.04-\gamma^c) x(t) + \gamma^c x^2(t) \bigr] \, \mathrm{mag}^2.
\label{eq:noise calculation appendix formula 2}
\end{equation}
The parameter $\gamma^c$ varies dependent on the filter from 0.037 to 0.040 and takes the value 0.039 for the $i$ band. Further,
$x=10^{0.4(m_\mathrm{AB,X}-m_5)}$, where $m_\mathrm{AB,X}$ is the AB magnitude in filter $X$ from Eq. (\ref{eq: microlensed light curves for ab magnitudes}) of the SN data
point and $m_5(t)$ is the 5$\sigma$ point-source depth (for more details see \cite{2009:LSSTscience}, Sec. 3.5, p. 67). The time varying 5$\sigma$ point-source depth comes from the moon phase as presented in \cite{Huber:2021iug}.

\end{document}

%% file: Time_delay_measurement_macro.tex
\newcommand{\bd}{\begin{displaymath}}
\newcommand{\ed}{\end{displaymath}}
\newcommand{\be}{\begin{equation}}
\newcommand{\ee}{\end{equation}}
\newcommand{\beaa}{\begin{eqnarray*}}
\newcommand{\eeaa}{\end{eqnarray*}}
\newcommand{\bea}{\begin{eqnarray}}
\newcommand{\eea}{\end{eqnarray}}

\newcommand{\erg}{\mathrm{erg}}

\newcommand{\cm}{\mathrm{cm}}

\newcommand{\sourcez}{z_{\rm s}}
\newcommand{\lensz}{z_{\rm d}}
\newcommand{\lum}{{D_\mathrm{lum}}}

\newcommand{\Rein}{R_\mathrm{Ein}}

\newcommand{\dd}{\mathrm{d}}

\DeclareSIUnit\parsec{pc}
\DeclareSIUnit\lightyear{ly}
\DeclareSIUnit\year{yr}
\DeclareSIUnit\erg{erg}
\DeclareSIUnit\ster{ster}
\DeclareSIUnit\arcsec{arcsec}
\DeclareSIUnit\rad{rad}
\DeclareSIUnit\mag{mag}

\usepackage{color} 
\definecolor{gruen}{cmyk}{0.35,0.01,0.80,0.1}

\definecolor{blue}{rgb}{0.0,0.0,1.0}
\definecolor{magenta}{rgb}{1.0,0.0,1.0}
\definecolor{orange}{rgb}{1.0,0.5,0.0}

%% file: LSTM_time_delay_SNIa.bbl
\begin{thebibliography}{104}
\expandafter\ifx\csname natexlab\endcsname\relax\def\natexlab#1{#1}\fi

\bibitem[{Abbott {et~al.}(2017)}]{LIGOScientific:2017adf}
Abbott, B.~P. {et~al.} 2017, Nature, 551, 85

\bibitem[{{Anand} {et~al.}(2024){Anand}, {Riess}, {Yuan}, {Beaton},
  {Casertano}, {Li}, {Makarov}, {Makarova}, {Tully}, {Anderson}, {Breuval},
  {Dolphin}, {Karachentsev}, {Macri}, \& {Scolnic}}]{2024arXiv240104776A}
{Anand}, G.~S., {Riess}, A.~G., {Yuan}, W., {et~al.} 2024, arXiv e-prints,
  arXiv:2401.04776

\bibitem[{Arendse {et~al.}(2023)Arendse, Dhawan, Carracedo, Peiris, Goobar,
  Wojtak, Alves, Biswas, Huber, \& Birrer}]{LSSTDarkEnergyScience:2023dbi}
Arendse, N., Dhawan, S., Carracedo, A.~S., {et~al.} 2023

\bibitem[{Bag {et~al.}(2021)Bag, Kim, Linder, \& Shafieloo}]{Bag_2021}
Bag, S., Kim, A.~G., Linder, E.~V., \& Shafieloo, A. 2021, \apj, 910, 65

\bibitem[{Balkenhol {et~al.}(2021)}]{SPT-3G:2021wgf}
Balkenhol, L. {et~al.} 2021, Phys. Rev. D, 104, 083509

\bibitem[{{Barnab{\`e}} {et~al.}(2011){Barnab{\`e}}, {Czoske}, {Koopmans},
  {Treu}, \& {Bolton}}]{Barnabe2011}
{Barnab{\`e}}, M., {Czoske}, O., {Koopmans}, L.~V.~E., {Treu}, T., \& {Bolton},
  A.~S. 2011, MNRAS, 415, 2215

\bibitem[{Bayer {et~al.}(2021)Bayer, Huber, Vogl, Suyu, Taubenberger, Sluse,
  Chan, \& Kerzendorf}]{Bayer:2021ugw}
Bayer, J., Huber, S., Vogl, C., {et~al.} 2021, A\&A, 653, A29

\bibitem[{{Bessell} \& {Murphy}(2012)}]{Bessel:2012}
{Bessell}, M. \& {Murphy}, S. 2012, \pasp, 124, 140

\bibitem[{Birrer {et~al.}(2019)}]{Birrer:2018vtm}
Birrer, S. {et~al.} 2019, MNRAS, 484, 4726

\bibitem[{Birrer {et~al.}(2020)}]{Birrer+2020}
Birrer, S. {et~al.} 2020, A\&A, 643, A165

\bibitem[{Blakeslee {et~al.}(2021)Blakeslee, Jensen, Ma, Milne, \&
  Greene}]{Blakeslee:2021rqi}
Blakeslee, J.~P., Jensen, J.~B., Ma, C.-P., Milne, P.~A., \& Greene, J.~E.
  2021, \apj, 911, 65

\bibitem[{Bonvin {et~al.}(2016)Bonvin, Tewes, Courbin, Kuntzer, Sluse, \&
  Meylan}]{Bonvin:2015jia}
Bonvin, V., Tewes, M., Courbin, F., {et~al.} 2016, A\&A, 585, A88

\bibitem[{Bonvin {et~al.}(2018)}]{Bonvin:2018dcc}
Bonvin, V. {et~al.} 2018, A\&A, 616, A183

\bibitem[{Breiman(2001)}]{breiman2001random}
Breiman, L. 2001, Mach. Learn., 45, 5

\bibitem[{Bromley {et~al.}(1993)Bromley, Bentz, Bottou, Guyon, Lecun, Moore,
  S\"{a}ckinger, \& Shah}]{doi:10.1142/S0218001493000339}
Bromley, J., Bentz, J.~W., Bottou, L., {et~al.} 1993, Int. J. Pattern Recognit.
  Artif. Intell., 07, 669

\bibitem[{{Chan} {et~al.}(2021){Chan}, {Rojas}, {Millon}, {Courbin}, {Bonvin},
  \& {Jauffret}}]{ChanEtal21}
{Chan}, J.~H.~H., {Rojas}, K., {Millon}, M., {et~al.} 2021, \aap, 647, A115

\bibitem[{Chen {et~al.}(2019)}]{Chen:2019ejq}
Chen, G. C.-F. {et~al.} 2019, \mnras, 490, 1743

\bibitem[{Courbin {et~al.}(2018)Courbin, Bonvin, Buckley-Geer, Fassnacht,
  Frieman, Lin, Marshall, Suyu, Treu, Anguita, \& et~al.}]{2017Courbin}
Courbin, F., Bonvin, V., Buckley-Geer, E., {et~al.} 2018, A\&A, 609, A71

\bibitem[{Cs\"ornyei {et~al.}(2023)Cs\"ornyei, Vogl, Taubenberger, Fl\"ors,
  Blondin, Cudmani, Holas, Kressierer, Leibundgut, \&
  Hillebrandt}]{Csornyei:2023rpw}
Cs\"ornyei, G., Vogl, C., Taubenberger, S., {et~al.} 2023, A\&A, 672, A129

\bibitem[{Denissenya \& Linder(2022)}]{Denissenya:2022eds}
Denissenya, M. \& Linder, E.~V. 2022, MNRAS, 515, 977

\bibitem[{Denzel {et~al.}(2021)Denzel, Coles, Saha, \&
  Williams}]{Denzel:2020zuq}
Denzel, P., Coles, J.~P., Saha, P., \& Williams, L. L.~R. 2021, MNRAS, 501, 784

\bibitem[{Dessart \& Hillier(2005)}]{Dessart:2005ax}
Dessart, L. \& Hillier, D.~J. 2005, A\&A, 437, 667

\bibitem[{Di~Valentino {et~al.}(2021)Di~Valentino, Mena, Pan, Visinelli, Yang,
  Melchiorri, Mota, Riess, \& Silk}]{DiValentino:2021izs}
Di~Valentino, E., Mena, O., Pan, S., {et~al.} 2021, Class. Quant. Grav., 38,
  153001

\bibitem[{Ding {et~al.}(2021)Ding, Liao, Birrer, Shajib, Treu, \&
  Yang}]{DingEtal21}
Ding, X., Liao, K., Birrer, S., {et~al.} 2021, MNRAS, 504, 5621

\bibitem[{Dutcher {et~al.}(2021)}]{SPT-3G:2021eoc}
Dutcher, D. {et~al.} 2021, Phys. Rev. D, 104, 022003

\bibitem[{{Falco} {et~al.}(1985){Falco}, {Gorenstein}, \&
  {Shapiro}}]{Falco:1985}
{Falco}, E.~E., {Gorenstein}, M.~V., \& {Shapiro}, I.~I. 1985, \apjl, 289, L1

\bibitem[{Foxley-Marrable {et~al.}(2018)Foxley-Marrable, Collett, Vernardos,
  Goldstein, \& Bacon}]{Foxley-Marrable:2018dzu}
Foxley-Marrable, M., Collett, T.~E., Vernardos, G., Goldstein, D.~A., \& Bacon,
  D. 2018, \mnras, 478, 5081

\bibitem[{Freedman(2021)}]{Freedman:2021ahq}
Freedman, W.~L. 2021, ApJ, 919, 16

\bibitem[{Freedman {et~al.}(2020)Freedman, Madore, Hoyt, Jang, Beaton, Lee,
  Monson, Neeley, \& Rich}]{Freedman_2020}
Freedman, W.~L., Madore, B.~F., Hoyt, T., {et~al.} 2020, ApJ, 891, 57

\bibitem[{Freedman {et~al.}(2019)}]{Freedman:2019jwv}
Freedman, W.~L. {et~al.} 2019, arXiv e-prints (arXiv:1907.05922)

\bibitem[{{Frye} {et~al.}(2024){Frye}, {Pascale}, {Pierel}, {Chen}, {Foo},
  {Leimbach}, {Garuda}, {Cohen}, {Kamieneski}, {Windhorst}, {Koekemoer},
  {Kelly}, {Summers}, {Engesser}, {Liu}, {Furtak}, {del Carmen Polletta},
  {Harrington}, {Willner}, {Diego}, {Jansen}, {Coe}, {Conselice}, {Dai},
  {Dole}, {D'Silva}, {Driver}, {Grogin}, {Marshall}, {Meena}, {Nonino},
  {Ortiz}, {Pirzkal}, {Robotham}, {Ryan}, {Strolger}, {Tompkins}, {Willmer},
  {Yan}, {Yun}, \& {Zitrin}}]{2024ApJ...961..171F}
{Frye}, B.~L., {Pascale}, M., {Pierel}, J., {et~al.} 2024, \apj, 961, 171

\bibitem[{Gayathri {et~al.}(2021)Gayathri, Healy, Lange, O'Brien, Szczepanczyk,
  Bartos, Campanelli, Klimenko, Lousto, \& O'Shaughnessy}]{Gayathri:2020mra}
Gayathri, V., Healy, J., Lange, J., {et~al.} 2021, ApJL, 908, L34

\bibitem[{{Goldstein} \& {Nugent}(2017)}]{GoldsteinNugent:2017}
{Goldstein}, D.~A. \& {Nugent}, P.~E. 2017, \apjl, 834, L5

\bibitem[{Goldstein {et~al.}(2018)Goldstein, Nugent, Kasen, \&
  Collett}]{Goldstein:2017bny}
Goldstein, D.~A., Nugent, P.~E., Kasen, D.~N., \& Collett, T.~E. 2018, ApJ,
  855, 22

\bibitem[{Goobar {et~al.}(2017)}]{Goobar:2016uuf}
Goobar, A. {et~al.} 2017, Science, 356, 291

\bibitem[{{Goobar} {et~al.}(2022){Goobar}, {Johansson}, {Dhawan}, {Schulze},
  {Arendse}, {Carracedo}, {Joseph}, {Nordin}, \&
  {Townsend}}]{2022TNSAN.180....1G}
{Goobar}, A.~A., {Johansson}, J., {Dhawan}, S., {et~al.} 2022, Transient Name
  Server AstroNote, 180, 1

\bibitem[{Hochreiter \& Schmidhuber(1997)}]{hochreiter1997long}
Hochreiter, S. \& Schmidhuber, J. 1997, Neural computation, 9, 1735

\bibitem[{Huber {et~al.}(2022)Huber, Suyu, Ghoshdastidar, Taubenberger, Bonvin,
  Chan, Kromer, Noebauer, Sim, \& Leal-Taix\'e}]{Huber:2021iug}
Huber, S., Suyu, S.~H., Ghoshdastidar, D., {et~al.} 2022, A\&A, 658, A157

\bibitem[{Huber {et~al.}(2021)Huber, Suyu, Noebauer, Chan, Kromer, Sim, Sluse,
  \& Taubenberger}]{Huber:2020dxc}
Huber, S., Suyu, S.~H., Noebauer, U.~M., {et~al.} 2021, A\&A, 646, A110

\bibitem[{Huber {et~al.}(2019)}]{Huber:2019ljb}
Huber, S. {et~al.} 2019, A\&A, 631, A161

\bibitem[{Ivezi\'c {et~al.}(2019)}]{Ivezic:2008fe}
Ivezi\'c, v. {et~al.} 2019, \apj, 873, 111

\bibitem[{Kasen {et~al.}(2006)Kasen, Thomas, \& Nugent}]{Kasen:2006ce}
Kasen, D., Thomas, R.~C., \& Nugent, P. 2006, ApJ, 651, 366

\bibitem[{{Kelly} {et~al.}(2022){Kelly}, {Zitrin}, {Oguri}, {Diego},
  {Williams}, {Broadhurst}, {Chen}, {Koekemoer}, {Pierel}, {Strolger}, \&
  {Treu}}]{2022TNSAN.169....1K}
{Kelly}, P., {Zitrin}, A., {Oguri}, M., {et~al.} 2022, Transient Name Server
  AstroNote, 169, 1

\bibitem[{Khetan {et~al.}(2021)}]{Khetan:2020hmh}
Khetan, N. {et~al.} 2021, A\&A, 647, A72

\bibitem[{{Kingma} \& {Ba}(2014)}]{Kingma:2014}
{Kingma}, D.~P. \& {Ba}, J. 2014, arXiv e-prints (arXiv:1412.6980)

\bibitem[{Kourkchi {et~al.}(2020)Kourkchi, Tully, Anand, Courtois, Dupuy,
  Neill, Rizzi, \& Seibert}]{Kourkchi:2020iyz}
Kourkchi, E., Tully, R.~B., Anand, G.~S., {et~al.} 2020, ApJ, 896, 3

\bibitem[{Kromer \& Sim(2009)}]{Kromer:2009ce}
Kromer, M. \& Sim. 2009, MNRAS, 398, 1809

\bibitem[{{Li} {et~al.}(2024){Li}, {Riess}, {Casertano}, {Anand}, {Scolnic},
  {Yuan}, {Breuval}, \& {Huang}}]{2024arXiv240104777L}
{Li}, S., {Riess}, A.~G., {Casertano}, S., {et~al.} 2024, arXiv e-prints,
  arXiv:2401.04777

\bibitem[{{Lochner} {et~al.}(2018){Lochner}, {Scolnic}, {Awan}, {Regnault},
  {Gris}, {Mandelbaum}, {Gawiser}, {Almoubayyed}, {Setzer}, {Huber}, {Graham},
  {Hlo{\v z}ek}, {Biswas}, {Eifler}, {Rothchild}, {Allam}, {Blazek}, {Chang},
  {Collett}, {Goobar}, {Hook}, {Jarvis}, {Jha}, {Kim}, {Marshall}, {McEwen},
  {Moniez}, {Newman}, {Peiris}, {Petrushevska}, {Rhodes}, {Sevilla-Noarbe},
  {Slosar}, {Suyu}, {Tyson}, \& {Yoachim}}]{Lochner:2018}
{Lochner}, M., {Scolnic}, D.~M., {Awan}, H., {et~al.} 2018, arXiv e-prints

\bibitem[{Lochner {et~al.}(2022)}]{LSSTDarkEnergyScience:2021ryz}
Lochner, M. {et~al.} 2022, ApJS, 259, 58

\bibitem[{{LSST Science Collaboration}(2009)}]{2009:LSSTscience}
{LSST Science Collaboration}. 2009, arXiv e-prints (arXiv:0912.0201)

\bibitem[{Maas {et~al.}(2013)Maas, Hannun, \& Ng}]{Maas:2013}
Maas, A.~L., Hannun, A.~Y., \& Ng, A.~Y. 2013, in ICML Workshop on Deep
  Learning for Audio, Speech and Language Processing

\bibitem[{Millon {et~al.}(2020)}]{Millon:2019slk}
Millon, M. {et~al.} 2020, A\&A, 639, A101

\bibitem[{Mueller \& Thyagarajan(2016)}]{Mueller_Thyagarajan_2016}
Mueller, J. \& Thyagarajan, A. 2016, Proceedings of the AAAI Conference on
  Artificial Intelligence, 30

\bibitem[{Mukherjee {et~al.}(2021)Mukherjee, Wandelt, Nissanke, \&
  Silvestri}]{PhysRevD.103.043520}
Mukherjee, S., Wandelt, B.~D., Nissanke, S.~M., \& Silvestri, A. 2021, Phys.
  Rev. D, 103, 043520

\bibitem[{Nix \& Weigend(1994)}]{374138}
Nix, D. \& Weigend, A. 1994, in Proceedings of 1994 IEEE International
  Conference on Neural Networks (ICNN'94), Vol.~1, 55--60 vol.1

\bibitem[{{Nomoto} {et~al.}(1984){Nomoto}, {Thielemann}, \&
  {Yokoi}}]{1984:Nomoto}
{Nomoto}, K., {Thielemann}, F.-K., \& {Yokoi}, K. 1984, ApJ, 286, 644

\bibitem[{{Oguri} \& {Kawano}(2003)}]{2003MNRAS.338L..25O}
{Oguri}, M. \& {Kawano}, Y. 2003, \mnras, 338, L25

\bibitem[{{Oguri} \& {Marshall}(2010)}]{Oguri:2010}
{Oguri}, M. \& {Marshall}, P.~J. 2010, MNRAS, 405, 2579

\bibitem[{{Pakmor} {et~al.}(2012){Pakmor}, {Kromer}, {Taubenberger}, {Sim},
  {R{\"o}pke}, \& {Hillebrandt}}]{Pakmor:2012}
{Pakmor}, R., {Kromer}, M., {Taubenberger}, S., {et~al.} 2012, \apjl, 747, L10

\bibitem[{Pascanu {et~al.}(2013)Pascanu, Mikolov, \& Bengio}]{Pascanu2013OnTD}
Pascanu, R., Mikolov, T., \& Bengio, Y. 2013, in ICML

\bibitem[{Paszke {et~al.}(2019)Paszke, Gross, Massa, Lerer, Bradbury, Chanan,
  Killeen, Lin, Gimelshein, Antiga, Desmaison, Kopf, Yang, DeVito, Raison,
  Tejani, Chilamkurthy, Steiner, Fang, Bai, \& Chintala}]{NEURIPS2019_9015}
Paszke, A., Gross, S., Massa, F., {et~al.} 2019, in Advances in Neural
  Information Processing Systems 32, ed. H.~Wallach, H.~Larochelle,
  A.~Beygelzimer, F.~Alche-Buc, E.~Fox, \& R.~Garnett (Curran Associates,
  Inc.), 8024--8035

\bibitem[{Pesce {et~al.}(2020)}]{Pesce:2020xfe}
Pesce, D.~W. {et~al.} 2020, ApJL, 891, L1

\bibitem[{Philipp {et~al.}(2018)Philipp, Song, \&
  Carbonell}]{Philipp2018GradientsE}
Philipp, G., Song, D.~X., \& Carbonell, J.~G. 2018, ArXiv, abs/1712.05577

\bibitem[{{Pierel} \& {Rodney}(2019)}]{PierelRodney+2019}
{Pierel}, J.~D.~R. \& {Rodney}, S. 2019, \apj, 876, 107

\bibitem[{{Planck Collaboration}(2020)}]{Planck:2018vks}
{Planck Collaboration}. 2020, A\&A., 641, A1

\bibitem[{Polletta {et~al.}(2023)}]{Polletta:2023vam}
Polletta, M. {et~al.} 2023, A\&A, 675, L4

\bibitem[{{Quimby} {et~al.}(2014){Quimby}, {Oguri}, {More}, {More}, {Moriya},
  {Werner}, {Tanaka}, {Folatelli}, {Bersten}, {Maeda}, \&
  {Nomoto}}]{Quimby:2014}
{Quimby}, R.~M., {Oguri}, M., {More}, A., {et~al.} 2014, Science, 344, 396

\bibitem[{{Refsdal}(1964)}]{Refsdal:1964}
{Refsdal}, S. 1964, MNRAS, 128, 307

\bibitem[{{Riess} {et~al.}(2024){Riess}, {Anand}, {Yuan}, {Casertano},
  {Dolphin}, {Macri}, {Breuval}, {Scolnic}, {Perrin}, \&
  {Anderson}}]{2024ApJ...962L..17R}
{Riess}, A.~G., {Anand}, G.~S., {Yuan}, W., {et~al.} 2024, \apjl, 962, L17

\bibitem[{Riess {et~al.}(2021)Riess, Casertano, Yuan, Bowers, Macri, Zinn, \&
  Scolnic}]{Riess:2020fzl}
Riess, A.~G., Casertano, S., Yuan, W., {et~al.} 2021, ApJL, 908, L6

\bibitem[{Riess {et~al.}(2019)Riess, Casertano, Yuan, Macri, \&
  Scolnic}]{Riess:2019cxk}
Riess, A.~G., Casertano, S., Yuan, W., Macri, L.~M., \& Scolnic, D. 2019, ApJ,
  876, 85

\bibitem[{{Riess} {et~al.}(2022){Riess}, {Yuan}, {Macri}, {Scolnic}, {Brout},
  {Casertano}, {Jones}, {Murakami}, {Anand}, {Breuval}, {Brink}, {Filippenko},
  {Hoffmann}, {Jha}, {D'arcy Kenworthy}, {Mackenty}, {Stahl}, \&
  {Zheng}}]{Riess+2022}
{Riess}, A.~G., {Yuan}, W., {Macri}, L.~M., {et~al.} 2022, \apjl, 934, L7

\bibitem[{Riess {et~al.}(2018)}]{Riess:2018byc}
Riess, A.~G. {et~al.} 2018, ApJ, 861, 126

\bibitem[{{Roberts-Pierel} \& {The LensWatch
  Collaboration}(2023)}]{2023AAS...24143207R}
{Roberts-Pierel}, J. \& {The LensWatch Collaboration}. 2023, in American
  Astronomical Society Meeting Abstracts, Vol.~55, American Astronomical
  Society Meeting Abstracts, 432.07

\bibitem[{Rodney {et~al.}(2021)Rodney, Brammer, Pierel, Richard, Toft,
  O'Connor, Akhshik, \& Whitaker}]{Rodney:2021keu}
Rodney, S.~A., Brammer, G.~B., Pierel, J. D.~R., {et~al.} 2021, arXiv e-prints
  (arXiv:2106.08935)

\bibitem[{{Rodney} {et~al.}(2015){Rodney}, {Patel}, {Scolnic}, {Foley},
  {Molino}, {Brammer}, {Jauzac}, {Brada{\v{c}}}, {Broadhurst}, {Coe}, {Diego},
  {Graur}, {Hjorth}, {Hoag}, {Jha}, {Johnson}, {Kelly}, {Lam}, {McCully},
  {Medezinski}, {Meneghetti}, {Merten}, {Richard}, {Riess}, {Sharon},
  {Strolger}, {Treu}, {Wang}, {Williams}, \& {Zitrin}}]{2015ApJ...811...70R}
{Rodney}, S.~A., {Patel}, B., {Scolnic}, D., {et~al.} 2015, \apj, 811, 70

\bibitem[{Rusu {et~al.}(2019)}]{Rusu:2019xrq}
Rusu, C.~E. {et~al.} 2019, arXiv e-prints (arXiv:1907.05922)

\bibitem[{Sak {et~al.}(2014)Sak, Senior, \& Beaufays}]{sak2014long}
Sak, H., Senior, A., \& Beaufays, F. 2014, arXiv preprint arXiv:1402.1128

\bibitem[{Saunders {et~al.}(2018)}]{Saunders:2018rjn}
Saunders, C. {et~al.} 2018, ApJ, 869, 167

\bibitem[{Schmidt {et~al.}(1998)}]{Schmidt:1998}
Schmidt, B.~P. {et~al.} 1998, ApJ, 507, 46

\bibitem[{Schneider \& Sluse(2014)}]{Schneider:2013wga}
Schneider, P. \& Sluse, D. 2014, A\&A, 564, A103

\bibitem[{{Seitenzahl} {et~al.}(2013){Seitenzahl}, {Ciaraldi-Schoolmann},
  {R{\"o}pke}, {Fink}, {Hillebrandt}, {Kromer}, {Pakmor}, {Ruiter}, {Sim}, \&
  {Taubenberger}}]{Seitenzahl:2013}
{Seitenzahl}, I.~R., {Ciaraldi-Schoolmann}, F., {R{\"o}pke}, F.~K., {et~al.}
  2013, \mnras, 429, 1156

\bibitem[{Shajib {et~al.}(2020)}]{Shajib:2019toy}
Shajib, A. {et~al.} 2020, MNRAS, 494, 6072

\bibitem[{{Shajib} {et~al.}(2018){Shajib}, {Treu}, \& {Agnello}}]{Shajib:2018}
{Shajib}, A.~J., {Treu}, T., \& {Agnello}, A. 2018, \mnras, 473, 210

\bibitem[{Sherstinsky(2020)}]{Sherstinsky_2020}
Sherstinsky, A. 2020, Physica D: Nonlinear Phenomena, 404, 132306

\bibitem[{{Sim} {et~al.}(2010){Sim}, {R{\"o}pke}, {Hillebrandt}, {Kromer},
  {Pakmor}, {Fink}, {Ruiter}, \& {Seitenzahl}}]{Sim:2010}
{Sim}, S.~A., {R{\"o}pke}, F.~K., {Hillebrandt}, W., {et~al.} 2010, \apjl, 714,
  L52

\bibitem[{{Sluse} {et~al.}(2019){Sluse}, {Rusu}, {Fassnacht}, {Sonnenfeld},
  {Richard}, {Auger}, {Coccato}, {Wong}, {Suyu}, {Treu}, {Agnello}, {Birrer},
  {Bonvin}, {Collett}, {Courbin}, {Hilbert}, {Koopmans}, {Tihhanova},
  {Marshall}, {Meylan}, {Shajib}, {Annis}, {Avila}, {Bertin}, {Brooks},
  {Buckley-Geer}, {Burke}, {Carnero Rosell}, {Carrasco Kind}, {Carretero},
  {Castander}, {da Costa}, {De Vicente}, {Desai}, {Doel}, {Evrard}, {Flaugher},
  {Frieman}, {Garc{\'\i}a-Bellido}, {Gerdes}, {Goldstein}, {Gruendl},
  {Gschwend}, {Hartley}, {Hollowood}, {Honscheid}, {James}, {Kim}, {Krause},
  {Kuehn}, {Kuropatkin}, {Lima}, {Lin}, {Maia}, {Marshall}, {Melchior},
  {Menanteau}, {Miquel}, {Plazas}, {Sanchez}, {Serrano}, {Sevilla-Noarbe},
  {Smith}, {Soares-Santos}, {Sobreira}, {Suchyta}, {Swanson}, \&
  {Tarle}}]{2019MNRAS.490..613S}
{Sluse}, D., {Rusu}, C.~E., {Fassnacht}, C.~D., {et~al.} 2019, \mnras, 490, 613

\bibitem[{Suyu {et~al.}(2017)}]{Suyu:2016qxx}
Suyu, S.~H. {et~al.} 2017, MNRAS, 468, 2590

\bibitem[{Suyu {et~al.}(2020)}]{Suyu:2020opl}
Suyu, S.~H. {et~al.} 2020, A\&A, 644, A162

\bibitem[{{Tewes} {et~al.}(2013){Tewes}, {Courbin}, \& {Meylan}}]{2013:Tewesb}
{Tewes}, M., {Courbin}, F., \& {Meylan}, G. 2013, A\&A, 553, A120

\bibitem[{Treu {et~al.}(2018)}]{DES:2018whv}
Treu, T. {et~al.} 2018, MNRAS, 481, 1041

\bibitem[{Vernardos \& Fluke(2014)}]{Vernardos:2014lna}
Vernardos, G. \& Fluke, C.~J. 2014, Astron. Comput., 6, 1

\bibitem[{Vernardos {et~al.}(2014)Vernardos, Fluke, Bate, \&
  Croton}]{Vernardos:2014yva}
Vernardos, G., Fluke, C.~J., Bate, N.~F., \& Croton, D. 2014, ApJS, 211, 16

\bibitem[{Vernardos {et~al.}(2015)Vernardos, Fluke, Bate, Croton, \&
  Vohl}]{Vernardos:2015wta}
Vernardos, G., Fluke, C.~J., Bate, N.~F., Croton, D., \& Vohl, D. 2015, ApJS,
  217, 23

\bibitem[{Vogl {et~al.}(2020)Vogl, Kerzendorf, Sim, Noebauer, Lietzau, \&
  Hillebrandt}]{Vogl:2019fhc}
Vogl, C., Kerzendorf, W.~E., Sim, S.~A., {et~al.} 2020, A\&A, 633, A88

\bibitem[{Vogl {et~al.}(2019)Vogl, Sim, Noebauer, Kerzendorf, \&
  Hillebrandt}]{Vogl:2018ckb}
Vogl, C., Sim, S.~A., Noebauer, U.~M., Kerzendorf, W.~E., \& Hillebrandt, W.
  2019, A\&A, 621, A29

\bibitem[{{Weisenbach} {et~al.}(2024){Weisenbach}, {Collett}, {Sainz de
  Murieta}, {Krawczyk}, {Vernardos}, {Enzi}, \&
  {Lundgren}}]{2024arXiv240303264W}
{Weisenbach}, L., {Collett}, T., {Sainz de Murieta}, A., {et~al.} 2024, arXiv
  e-prints, arXiv:2403.03264

\bibitem[{{Weisenbach} {et~al.}(2021){Weisenbach}, {Schechter}, \&
  {Pontula}}]{Weisenbach+2021}
{Weisenbach}, L., {Schechter}, P.~L., \& {Pontula}, S. 2021, \apj, 922, 70

\bibitem[{Wojtak {et~al.}(2019)Wojtak, Hjorth, \& Gall}]{Wojtak:2019hsc}
Wojtak, R., Hjorth, J., \& Gall, C. 2019, MNRAS, 487, 3342

\bibitem[{Wong {et~al.}(2020)}]{Wong:2019kwg}
Wong, K.~C. {et~al.} 2020, \mnras, 498, 1420

\bibitem[{Yahalomi {et~al.}(2017)Yahalomi, Schechter, \&
  Wambsganss}]{Yahalomi:2017ihe}
Yahalomi, D.~A., Schechter, P.~L., \& Wambsganss, J. 2017, arXiv e-prints
  (arXiv:1711.07919)

\bibitem[{{Y{\i}ld{\i}r{\i}m} {et~al.}(2020){Y{\i}ld{\i}r{\i}m}, Suyu, \&
  Halkola}]{Yildirim:2019vlv}
{Y{\i}ld{\i}r{\i}m}, A., Suyu, S.~H., \& Halkola, A. 2020, MNRAS, 493, 4783

\bibitem[{{Y{\i}ld{\i}r{\i}m} {et~al.}(2017){Y{\i}ld{\i}r{\i}m}, {van den
  Bosch}, {van de Ven}, {Mart{\'{\i}}n-Navarro}, {Walsh}, {Husemann},
  {G{\"u}ltekin}, \& {Gebhardt}}]{2017:Yildirim}
{Y{\i}ld{\i}r{\i}m}, A., {van den Bosch}, R.~C.~E., {van de Ven}, G., {et~al.}
  2017, MNRAS, 468, 4216

\end{thebibliography}
